\documentclass[preprint,review,1p]{elsarticle}

\usepackage{graphicx}
\usepackage{newtxtext}
\usepackage{newtxmath}
\usepackage{natbib}
\usepackage{hyperref}
\hypersetup{
    colorlinks = true,
    urlcolor   = blue,
    citecolor  = blue,
}

\newcommand{\RomanNumeralCaps}[1]
\linenumbers

\usepackage{enumitem}
\usepackage{caption}
\usepackage{subcaption}
\usepackage{graphicx}
\usepackage{placeins}
\usepackage{dcolumn}

\usepackage{hyperref}
\usepackage[utf8]{inputenc}
\usepackage[T1]{fontenc}
\usepackage{etoolbox}
\usepackage{algorithm,algpseudocode}
\usepackage{booktabs}
\usepackage{bm} 

\newcommand{\vect}[1]{\bm{#1}}

\newcommand{\matr}[1]{\bm{\mathsf{#1}}}

\journal{Computer Methods in Applied Mechanics and Engineering}

\begin{document}

\begin{frontmatter}

\title{Real-time forecasting of chaotic dynamics from sparse data and autoencoders}

\author[label1]{Elise \"Ozalp}
\author[label1]{Andrea Nóvoa}
\author[label1,label2,label3]{Luca Magri\corref{cor1}}
\cortext[cor1]{Corresponding author: l.magri@imperial.ac.uk}

\affiliation[label1]{organization={Department of Aeronautics, Imperial College London},
            addressline={South Kensington Campus},
            city={London},
            postcode={SW7 2BX},
            country={United Kingdom}}

\affiliation[label2]{organization={DIMEAS, Politecnico di Torino},
            addressline={Corso Duca degli Abruzzi, 24},
            city={Torino},
            postcode={10129},
            country={Italy}}

\affiliation[label3]{organization={The Alan Turing Institute},
            addressline={British Library, 96 Euston Road},
            city={London},
            postcode={NW1 2DB},
            country={United Kingdom}}



\begin{abstract}

The real-time prediction of chaotic systems requires a nonlinear-reduced order model (ROM) to forecast the dynamics, and a stream of data from sensors to update the ROM. Data-driven ROMs are typically built with a two-step strategy: data compression in a lower-dimensional latent space, and prediction of the temporal dynamics on it. To achieve real-time prediction, however, there are two challenges to overcome: 
(i) ROMs of chaotic systems can become numerically unstable; and 
(ii) sensors’ data are sparse, i.e., partial, and noisy. 
To overcome these challenges, we propose a three-step strategy: 
(i) a convolutional autoencoder (CAE) compresses the system's state onto a lower-dimensional latent space;  
(ii) a latent ROM  (echo state network, ESN), which is formulated as a state-space model, predicts the temporal evolution on the latent space; and 
(iii) sequential data assimilation based on the Ensemble Kalman filter (EnKF) adaptively corrects the latent ROM by assimilating noisy and sparse measurements. This provides a numerically stable method (DA-CAE-ESN), which corrects itself every time that data becomes available from sensors. 
The DA-CAE-ESN is tested on spatio-temporally chaotic partial differential equations: the Kuramoto–Sivashinsky equation, and a two-dimensional Navier-Stokes equation (Kolmogorov flow). We show that the method provides accurate and stable forecasts across different levels of noise, sparsity, and sampling rates. As a by-product, the DA-CAE-ESN acts as a localization strategy that mitigates spurious correlations, which arise when applying the EnKF to high-dimensional systems.  The DA-CAE-ESN provides a numerically stable method to perform real-time predictions, which opens opportunities for deploying data-driven latent models.

\end{abstract}


\end{frontmatter}


\section{Introduction}\label{sec:intro}
Spatio-temporally chaotic systems  are central in a wide range of problems, from weather~\cite{lorenz63, mass1998regional, cassola2012wind, robinson1984real, lermusiaux2006uncertainty} to aerodynamics~\cite{lee1999nonlinear, kato2015data}, propulsion systems~\cite{huhn2020stability}, and turbulence~\cite{holmes2012turbulence}, among many others. 
%
%
Forecasting the dynamics of spatio-temporally chaotic systems is challenging because 
(i) the state evolves in a high-dimensional space~\cite{foias2001navier}; and  
(ii) infinitesimal perturbations grow exponentially both in space and time~\cite{alligood1997chaos}.
The overarching objective of this paper is to develop an adaptive nonlinear-reduced-order model, which forecasts the spatio-temporal dynamics as data from sensors become available (real-time). 
This is achieved by a three-step strategy:  
(i) compression of the high-dimensional dynamics into a latent space;
(ii) prediction of the temporal dynamics on the latent space; 
and 
(iii) a data assimilation method, which assimilates sparse and noisy data from the physical space to the latent space in real-time. \\ 

Although spatio-temporal chaos is high-dimensional, its long-term behaviour  evolves on a lower-dimensional subset of the phase space, which is known as the inertial manifold~\cite{foias1988inertial,hilborn2000chaos}.
The availability of high-fidelity experimental and numerical data has motivated the development of data-driven reduced-order models (ROMs) to extract interpretable representations, for example, of fluid dynamic systems~\citep[e.g.,][]{taira2017modal, duraisamy2019turbulence, mendez2023data}.  
%
%
Linear techniques, such as proper orthogonal decomposition (POD)~\cite{lumley1970toward} and dynamic mode decomposition~\cite{Schmid2010}, have been adopted to identify dominant coherent structures and enable reduced-order modelling. However, these techniques often require a large number of modes to capture  nonlinear dynamics~\cite[e.g.,][]{alfonsi2007structure, osth2014need, noack2016recursive, muralidhar2019spatio, rozza2022advanced}. 
In contrast, nonlinear approaches, e.g., convolutional autoencoders (CAEs), can efficiently encode high-dimensional flow fields into compact latent representations~\citep[e.g.,][]{uribarri2020structure, fukami2020convolutional, de2023data, Mo_Traverso_Magri_2024}, which capture physically meaningful properties of the system~\citep[e.g.,][]{page2021revealing, ozalp2025stability}. 
Once a low-dimensional latent representation is constructed, we can complete the ROM by learning the temporal evolution of the system in the latent coordinates using 
long-short term memory networks (LSTMs)~\cite{Hochreiter_1997_LongShortTermMemory,nakamura2021convolutional,borrelli2022predicting, vlachas2022multiscale, corrochano2025predictive}, 
reservoir computers~\cite{jaeger2001echo,Vlachas_2020_backprop,racca2023predicting}, transformers~\cite{vaswani2017attention,solera2024beta, shokar2024stochastic, mentzelopoulos2024variational} or 
neural ordinary differential equations~\cite{chen2018neural,linot2022data, linot2023stabilized, de2023data}, amongst other architectures~\cite{herzog2019convolutional, maulik2021reduced, gupta2023mori}. 
Unlike models that require iterative backpropagation, echo state networks (ESNs)~\cite{Vlachas_2020_backprop,racca2023predicting} are trained by solving one quadratic optimization problem in the latent space, and naturally provide a state-space formulation~\cite{Magri2024vki}.
The strategy of compressing the state in the latent space and predicting its temporal evolution has shown success for different nonlinear problems~\cite{hasegawa2020cnn, linot2020deep, linot2022data, vlachas2022multiscale, gupta2023mori,racca2023predicting}. 
One problem is still open: When used for autonomous forecasting in chaotic systems, autoencoder-based models may become numerically unstable and drift toward unphysical solutions\footnote{This problem was also stated in a keynote talk "AI  and Scientific Computing: there is plenty of room in the middle" by Prof. Petros Koumoutsakos (Harvard) in the Joint event
"Euromech Colloquium on Data-Driven Fluid Dynamics"
\&
"2nd ERCOFTAC Workshop on Machine Learning for Fluid Dynamics", London, 2-4 April 2025.\label{fr3frrfed44}}~\citep[e.g.,][]{linot2022data,vlachas2022multiscale, racca2023predicting, ozalp2025stability}. 
To mitigate this in neural ordinary differential equations, \cite{linot2023stabilized} introduced an damping term to suppress high-wavenumber modes, however, this is an ad-hoc strategy. In this paper, we offer a solution to this open problem by combining data assimilation into  latent models to stabilize their long-term evolution. \\

In practice, predictions can be improved by incorporating real-time measurements (observations) of the system.  Sequential data assimilation methods, such as the Kalman filter~\cite{kalman1960new}, address this by  combining model forecasts with observations to perform online state and parameter estimation and to quantify the model uncertainty~\cite{tarantola_inverse_2005, evensen2009data}. 
The Kalman filter is a Bayesian filter that solves the inverse problem from a maximum a posteriori approach. 
The ensemble Kalman filter (EnKF) ~\cite{evensen2003ensemble} enables the application of the Bayesian framework to nonlinear dynamical systems and turbulence~\citep[e.g.,][]{da2018ensemble, labahn2020ensemble, spratt2021characterizing, zhang2022ensemble, ZAKI2025129}. 
%
Ensemble filters have been used to perform real-time state estimation in low-dimensional systems~\citep[e.g.,][]{reich2015probabilistic, darakananda2018data, goswami2021data, novoa2022real,novoa2024inferring}. 
Although these filters perform well in capturing system dynamics and uncertainty for low-dimensional problems, their application to high-dimensional settings presents challenges, e.g., spurious correlations.  A comprehensive review of methods, which aim to bridge data-driven ROM and data assimilation, is provided by \cite{BUIZZA2022101525, cheng_2023_mldauqreview}.
Broadly, existing approaches fall into two categories: (i) full projection and (ii) partial projection methods. Full projection methods assume that the full state can be measured at each assimilation step, which enables the assimilation in the latent space after encoding the full-state observation~\cite{amendola2021data, peyron2021latent, HEANEY2024129783}.  
In practice, however, only sparse or partial observations are available.  Partial projection methods relax the full-observability assumption by applying the full-state encoder to partial observations, which is achieved by projecting the observations onto a reduced basis (e.g., POD)~\cite{HEANEY2024129783} or via preprocessing techniques (e.g., masking or inpainting)~\cite{cheng2022data, cheng2023generalised, cheng2024multidomain}.  
The choice of the reduced basis or preprocessing strategy, however,  is an approximation, which influences the assimilation accuracy and computational efficiency. When observations are significantly sparse or differ in distribution from the training data, partial encodings may not provide robust results. Therefore, assimilating sparse and noisy observations  into  latent ROMs remains an open challenge~\cite[e.g.,][]{dee_data_1998, drecourt_bias_2006, farchi_using_2021}. \\

In this work, we propose a   data assimilation framework for real-time forecasting of high-dimensional spatio-temporally chaotic systems. The framework combines
(i) a convolutional autoencoder (CAE) to compute the latent space;
(ii)  an echo state network (ESN) to model the  temporal dynamics in the latent space; 
and 
(iii) an ensemble Kalman filter (EnKF) to assimilate sparse and noisy data. 
The tool's acronym is DA-CAE-ESN for brevity. 
Section~\ref{sec:cae-esn} introduces  the CAE-ESN. 
Section~\ref{sec:data-assimilation} provides the theoretical background and implementation of the EnKF in the latent space with an augmented state-space formulation.
The DA-CAE-ESN is applied to the Kuramoto-Sivashinsky (KS) equation and Kolmogorov flow, whose results are reported in Sections~\ref{sec:ks-results} and~\ref{sec:results-kolmogorov}, respectively.
Section~\ref{sec:conclusion} concludes the work.

\section{Learning latent dynamics}\label{sec:cae-esn}

Consider a nonlinear dynamical system evolving the state $\vect{u}(\vect{x}, t)\in \mathbb{R}^N$ according to 
\begin{align}\label{eq:dyn_system}
    \frac{d}{dt} \vect{u}(\vect{x}, t) = \mathcal{F}\left(\vect{u}(\vect{x}, t)\right),
\end{align}
where  $\vect{x}$ is defined on the domain $\partial D$, and the operator $\mathcal{F}: \mathbb{R}^N \to \mathbb{R}^N$ is a smooth and nonlinear function, which, in fluid mechanics represents the Navier-Stokes operator.  
Direct numerical simulation of such systems becomes computationally prohibitive due to the high dimensionality of $N$, which scales with the spatial resolution of $\vect{x}$. 
Following prior work on reduced-order modelling of chaotic systems~\cite{heyder2022generalizability, vlachas2022multiscale, racca2023predicting,ozalp2025stability}, we adopt a hybrid architecture that combines a CAE with an ESN. In summary, the CAE compresses the high-dimensional flow data onto a nonlinear latent manifold, capturing key spatial features~\cite{Mo_Traverso_Magri_2024}, while the ESN models the temporal evolution of the low-dimensional latent variables. The two components are trained separately, as end-to-end optimization has not demonstrated superior forecasting accuracy and incurs higher computational cost~\cite{vlachas2022multiscale}. 
The overall structure of the CAE-ESN is illustrated Fig.~\ref{fig:schematic}. 
On the one hand, 
autoencoders~\cite{hinton1993autoencoders, vincent2008extracting} are based on neural networks that learn efficient lower-dimensional encodings of high-dimensional input data, capturing the main features of the input~\cite{wang2014generalized, fefferman2016testing}. In chaotic  systems, such as the Kuramoto–Sivashinsky equation and the Kolmogorov flow, well-trained autoencoders have been shown to preserve key stability characteristics of the original dynamics~\cite{page2021revealing, ozalp2025stability}.
In this work, the CAE maps each flow snapshot to a compact latent representation, extracting essential spatial features, as detailed in \S\ref{subsec:cae}. This process yields a low-dimensional time series, which is the input to the temporal evolution model. 
On the other hand, 
ESNs are a class of reservoir computing models suited for modelling chaotic fluid dynamics~\citep[e.g.,][]{huhn2020stability, racca2021robust}. Their efficiency is enabled by a training procedure based on solving a linear least-squares problem, avoiding back-propagation. ESNs also naturally yield a nonlinear state-space formulation~\cite{Magri2024vki}, which facilitates data assimilation (\S\ref{sec:data-assimilation}) and enables data-driven stability analysis~\cite{Pathak_2017_ml_le, Margazoglou2023stability, ozalp2025stability}. 
Section~\ref{subsec:esn} details the ESN implementation  following established best practices and recent advances in ESN modelling~\cite{lukovsevivcius2012practical, racca2021robust, Margazoglou2023stability, racca2023predicting, ozalp2025stability}.

\subsection{Data compression with a convolutional autoencoder}\label{subsec:cae}

 An autoencoder approximates an identity mapping through two networks: an encoder and a decoder. The encoder  $\mathcal{E}$ maps the physical state $\vect{u}(t_k)$ to a latent state $ \vect{y}(t_k) \in \mathbb{R}^{N_\text{lat}}$ on the manifold, where $N_\text{lat} \ll N$. The decoder $\mathcal{D}$ reconstructs the physical state from the latent representation by mapping back to the full state, 
\begin{align}\label{eq:AE}
    \vect{y}(t_k)=\mathcal{E}\left(\vect{u}(t_k) \right), 
     \; \text{and }\;
      \mathcal{D}\left(\vect{y}(t_k)\right)
      =\vect{\hat{u}}(t_k)\approx \vect{u}(t_k),  
\end{align}
where $\vect{\hat{u}}(t_k)$ is output of the autoencoder. 
We use convolutional layers~\cite{lecun1989generalization} in both $\mathcal{E}$ and $\mathcal{D}$ to capture the spatial structure of the input $\vect{u}(t_k)$, resulting in a CAE. 
Each convolutional layer is followed by a $\tanh(\cdot)$ activation function, as linear activation functions make the network equivalent to POD~\cite{brunton2020machine, Mo_Traverso_Magri_2024}. 
For flow fields, we adopt a multi-scale CAE~\cite{du2018single}, which has proven successful in generating latent representations for  flow fields~\cite{nakamura2021convolutional, racca2023predicting}. 
The multi-scale CAE consists of three parallel encoders and decoders, but with a shared latent manifold. Each encoder-decoder pair is characterised by a different kernel size to learn spatial structures of different sizes, which are characteristic features of flows. 
The CAE is trained to minimize the mean squared error (MSE)
\begin{align}
    \mathcal{L}\left(\vect{u}, \vect{\hat{u}}\right) = \frac{1}{N_\text{train}} \sum_{k=1}^{N_\text{train}} \| \vect{u}(t_k) - \vect{\hat{u}}(t_k) \|_2^2, 
\end{align}
where $N_\text{train}$ is the number of training snapshots. Once trained, the autoencoder encodes the physical solution $\vect{u}(t_k)$ to the latent representation $\vect{y}(t_k)$ of it, resulting in the latent training data $\{ \vect{y}(t_k)\}^{N_\text{train}}_{k=1}$. 
Further details on the implementation can be found in \ref{app:cae-esn-details}, and for a detailed discussion on selecting an appropriate latent space dimension for chaotic systems, we refer to~\cite{vlachas2022multiscale, racca2023predicting, ozalp2025stability}.

\subsection{Temporal prediction with an echo state network}\label{subsec:esn}

Echo state networks (ESNs) are designed to learn the temporal dynamics of sequential data through a hidden high-dimensional reservoir state $\vect{r}(t_k)$.  
The time evolution of the reservoir state can be expressed in the state-space formulation, as explained by~\cite{Magri2024vki}
\begin{align}\label{eq:esn1}
        \vect{r}(t_{k+1}) &= \tanh\left([\vect{y}_\text{in}(t_k); b_\text{in}]^\top\matr{W}_\text{in} + \vect{r}^\top(t_k)\matr{W} \right). 
\end{align}
where
$\vect{y}_\text{in}$ is the input to the network, which is augmented with the input bias $b_\text{in}$ to break the symmetry of $\tanh$, and $[~;~]$ indicates vertical concatenation.  
The input is mapped to the reservoir via the input matrix $\matr{W}_\text{in} \in \mathbb{R}^{N_\text{lat}+1 \times N_r}$, which is a dense matrix whose entries are sampled from a uniform distribution $\mathcal{U}[-\sigma_\text{in}, \sigma_\text{in}]$ with input scaling $\sigma_\text{in}$. 
%
The reservoir state $\vect{r}(t_{k+1})$ retains information from the previous reservoir state $\vect{r}(t_k)$ through the state matrix $\matr{W} \in \mathbb{R}^{N_r \times N_r}$, which is an Erd\"os-Renyi sparse matrix with connectivity $c$, i.e.,  each row of $\matr{W}$ has only $c$ non-zero elements sampled from $\mathcal{U}[-1, 1]$ and subsequently rescaled to a set spectral radius $\rho$. 
The predicted latent state $\vect{\Hat{y}}(t_{k+1})$ is computed through the linear readout  
\begin{align}\label{eq:esn2}
  \vect{\hat{y}}(t_{k+1}) & = \vect{r}^\top(t_{k+1}) \matr{W}_\text{out}, 
\end{align}
where 
$\matr{W}_\text{out} \in \mathbb{R}^{N_r \times N_\text{lat}}$ is the output matrix, which is the only trainable part of the ESN.  
Training $\matr{W}_\text{out}$ consists of minimising the MSE between the network prediction $ \vect{\Hat{y}}(t_k)$ and the reference data $\vect{y}(t_k)$ over the training dataset
\begin{align}\label{eq:loss_esn}
    \mathcal{L}\left(\vect{y}_\text{in}, \vect{\hat{y}}\right) = \frac{1}{N_\text{train}} \sum_{k=1}^{N_\text{train}} \| \vect{y}_\text{in}(t_k) - \vect{\hat{y}}(t_k) \|_2^2.
\end{align}
Eq.~\eqref{eq:loss_esn} is a quadratic optimisation problem. Instead of backpropagation, the optimal solution is obtained by solving one ridge regression system 
\begin{align}
    (\matr{R}\matr{R}^\top + \beta \mathbb{I}_{N_r}) \matr{W}_\text{out} = \matr{R}{\matr{Y}}^\top,
\end{align}
where 
$\matr{Y}=[\vect{y}(t_2),\dots,\vect{y}(t_{N_\text{train}})]$ is the training data, 
$\matr{R} = [\vect{r}(t_2), \dots, \vect{r}(t_{N_\text{train}})]$ is computed via Eq.~\eqref{eq:esn1} using the training data $\matr{Y}_\text{in}=[\vect{y}(t_1),\dots,\vect{y}(t_{N_\text{train}-1})]$ as input, 
$\mathbb{I}_{N_r}$ is the identity matrix of size $N_r$, 
and $\beta$ is the Tikhonov regularization parameter. The hyperparameters $\{\beta$, $\rho$, $\sigma_\text{in}\}$ are optimized with a grid search followed by a Bayesian optimisation to maximize short-term prediction on the different intervals of the validation data (see ~\ref{app:cae-esn-details}).
\begin{figure}
    \includegraphics[width=\textwidth]{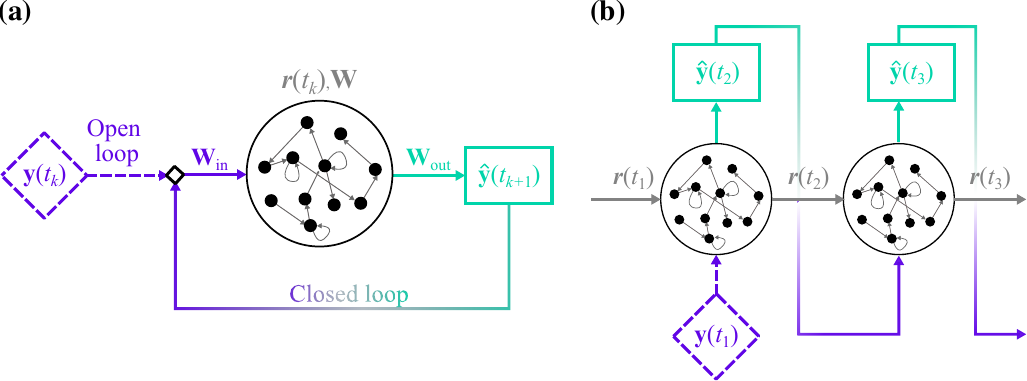}
    \caption{Echo state network architecture. (a) Compact representation of the open-loop and closed-loop mode based on the network input and (b) unfolded representation illustrating closed-loop steps.} \label{fig:esn_open_closed_loop}
\end{figure} 

The reservoir can operate in two configurations: open-loop and closed-loop, which differ in $\vect{y}_\text{in}$, as depicted in Fig.~\ref{fig:esn_open_closed_loop}.   
When input data is available, the network propagates in open-loop, such that $\vect{y}_\text{in}(t_k)=\vect{y}(t_k)$, which is an encoded flow field snapshot.  
The network runs in open-loop during training and during the initialization phase, i.e., the washout (see \S\ref{sec:implement} for details). 
Once trained, the ESN runs autonomously in closed-loop, such that the predicted latent state becomes the input to the next step, i.e., $\vect{y}_\text{in}(t_k)=\vect{\hat{y}}(t_k)$, where $\vect{\hat{y}}$ is forecasted via Eq.~\eqref{eq:esn2}. To obtain the full state forecast, the latent prediction is decoded as $\vect{\hat{u}}(t_k) = \mathcal{D}(\vect{\hat{y}}(t_k))$.

\subsection{Discussion}\label{sec:discussion}
By construction, there are limitations to the CAE-ESN architecture. The encoder requires the full system snapshot to map the physical state into the latent space. 
Once the input has been encoded, the ESN evolves autonomously, and noise is propagated through the autonomous forecast. New information can only be incorporated through the open-loop mode, when new latent inputs are provided.  
The CAE-ESN model described in \S\ref{sec:cae-esn}  offers a computationally efficient approach to forecasting high-dimensional dynamical systems and yields accurate and stable short-term predictions across quasi-periodic and chaotic regimes~\citep[e.g.,][]{ vlachas2022multiscale, racca2023predicting, ozalp2025stability}. 
%
%
Long-term autonomous forecasting, particularly in chaotic regimes, often becomes numerically unstable in the long term~\cite{vlachas2022multiscale, racca2023predicting, ozalp2025stability}, which is exacerbated when data is noisy. 
The onset of instabilities appears to depend on a combination of factors, including the network architecture, the network initialization, and the complexity of the system’s attractor. These effects are likely network- and system-specific, as evidenced by observations across different studies in the literature~\cite{dueben2018challenges, vlachas2018data, linot2023stabilized, racca2023predicting} (see also footnote~\ref{fr3frrfed44}). 
%
In Section~\ref{sec:data-assimilation}, we propose a solution by introducing the data-assimilated CAE-ESN (DA-CAE-ESN).

\section{Real-time modelling via data assimilation}\label{sec:data-assimilation}

\begin{figure}[!htb]
    \centering
    \includegraphics[width=\linewidth]{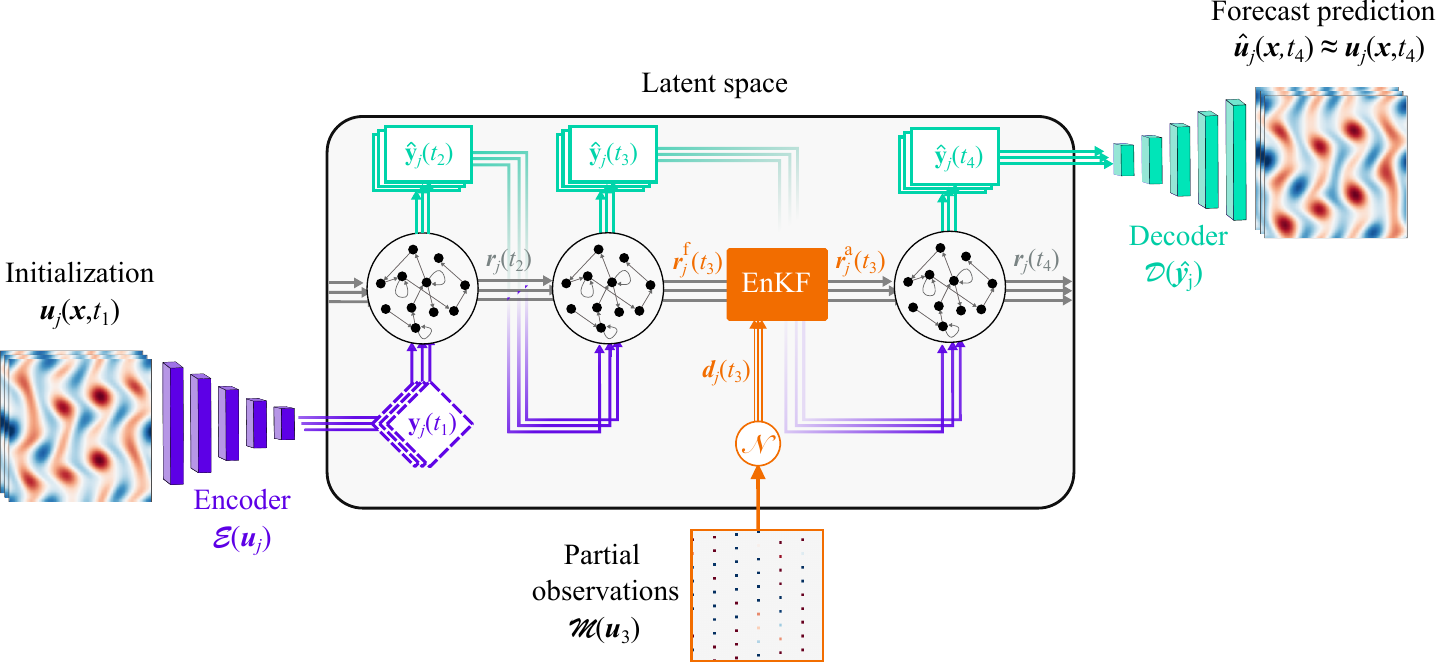}
    \caption{Illustration of the proposed DA-CAE-ESN framework, which integrates a convolutional autoencoder, an echo state network and an ensemble Kalman filter with Kolmogorov flow data.  The CAE encodes an ensemble of initial snapshots into a latent state $\vect{y}$ at a time $t_1$ via the encoder $\mathcal{E}$. 
    In the latent space, the ESN forecasts in closed-loop an ensemble of predicted latent states $\vect{\hat{y}}(t_j)$ until observations become available (here, for illustration at $t=t_3$). Then, the EnKF combines the partial and noisy observations $\vect{d}$ with the ensemble of reservoir states $\vect{r}_j^f$, resulting in the analysis ensemble $\vect{r}^a(t_j)$. The updated ESN can continue to forecast in closed-loop. The decoder $\mathcal{D}$ recovers the prediction of the full flow field $\vect{\hat{u}}(t_j)$ from any $\vect{\hat{y}}(t_j)$ (here, for illustration at $t=t_4$). }
    \label{fig:schematic}
\end{figure}

We deploy a sequential data assimilation method to correct the forecast using observations, which may be noisy and sparse. Specifically, we focus on the Ensemble Kalman Filter (EnKF) because its sequential nature and Monte Carlo approximation make it well-suited for real-time forecasting~\cite{evensen2003ensemble, darakananda2018data, novoa2022real}. The proposed methodology is pictorially illustrated in Figure~\ref{fig:schematic}.  
Sequential data assimilation consists of iteratively repeating the following steps: 
(i) the forecast step, in which the model evolves the state forward in time, and 
(ii) the analysis step, in which new measurements are assimilated to update the state estimate.  
The updated state estimate is the new initial condition for the next forecast step.

We formulate the sequential data assimilation problem from a Bayesian perspective (\S\ref{sec:da-problem-statement}) and present the ensemble Kalman filter (\S\ref{sec:enKF}). 
 We define a general framework that, albeit demonstrated on the CAE-ESN, applies broadly to other data-driven ROMs with recurrent components. The approach operates directly on both the model output and the recurrent states, making it easily adaptable to other architectures such as the POD-ESN \cite{novoa_online_2025}, autoencoder–LSTM~\cite{vlachas2022multiscale}, or variational autoencoder-transformer~\cite{solera2024beta, mentzelopoulos2024variational}.

\subsection{Data assimilation for reduced order models}\label{sec:da-problem-statement}

In data-driven reduced order modelling, we approximate~\eqref{eq:dyn_system} as
\begin{subequations}\label{eq:ROM-DA}
\begin{align}
   \vect{r}(t_{k+1}) &= f_1(\vect{r}(t_k))\\ 
    \vect{\hat{u}}(\vect{x}, t_{k+1}) &= f_2(\vect{r}(t_{k+1})) + \vect{\epsilon}_f,
\end{align}
\end{subequations}
where 
 $\vect{r} \in \mathbb{R}^{N_r}$ is the state vector of the ROM, 
 $f_1$ is the surrogate model that forecasts the ROM state vector approximating the true system dynamics, and
 $f_2$ is the decoding that maps the ROM state to the physical state. The term $\vect{\epsilon}_f$ represents the error introduced with the ROM approximation (i.e., by $f_2$) and by discretizing the system in time $t_k$ and in space $\vect{x} = [x_1; \dots; x_N]$.  
Specifically, for the CAE-ESN model deployed in this work,  
 $\vect{r}$ is the ESN reservoir state, 
 $f_1$ is the reservoir Eq.~\eqref{eq:esn1}, and 
 the decoding function is $f_2\left(\vect{r}\right) = \mathcal{D}\left(\vect{r}^\top\matr{W}_\text{out}\right)$ from Eqs.~\eqref{eq:AE}~and~\eqref{eq:esn2}.

As discussed in \S\ref{sec:intro},  existing methods combining data assimilation and reduced order models either assume access to full-state observations or rely on data preprocessing techniques to encode partial measurements. However, in practice, especially for high-dimensional and chaotic systems, only sparse and spatially limited observations are available. 
To address this challenge, we propose a method to correct the latent state directly using an augmented state-space formulation~\citep[e.g.,][]{novoa2022real,novoa2024real, ozan2025data}. 
Let $\mathcal{M}: \mathbb{R}^N \to \mathbb{R}^{N_\text{obs}}$ denote a measurement operator that maps the full state to observable quantities, yielding measurements of the form 
\begin{equation}
    \vect{d}(\vect{x},t_k) = \mathcal{M}\left(\vect{u}(\vect{x}, t_k)\right) + \vect{\epsilon}_d.
\end{equation}
where $\vect{\epsilon}_d$ is measurement noise, 
and $N_\text{obs} \ll N$ reflects the limited number of observation points. 
We define the augmented state vector $\vect{\psi} = \left[\vect{r};\mathcal{M}\left(\vect{\hat{u}}\right)\right]\in \mathbb{R}^{N_\psi=N_r+N_\text{obs}}$, such that the  measurement operator in augmented state-space $\matr{M} = \left[\matr{0}_{N_\text{obs}\times N_r}; \mathbb{I}_{N_\text{obs}} \right]$ is linear and  the model prediction of the measurements is 
$\matr{M}\vect{\psi}=\mathcal{M}\left(\vect{\hat{u}}\right)$. 
This augmented state-space approach effectively 
facilitates the assimilation of partial measurements through the decoder~\cite{novoa_online_2025,ozan2025data}.  
\\

Our objective is to exploit the information from the sparse sensors to update the predictions from the ROM in real time.  
Here, we update the reservoir state of the ESN as this yields faster convergence and accuracy than updating the ESN's prediction on the latent state~\cite{novoa_online_2025}. 
 %
%
To find the \textit{analysis} state, i.e., the state that best captures the evolution of the system given the surrogate model $\vect{F}=[f_1; f_2]$ and the noisy, partial observations $\vect{d}$, we seek a Bayesian maximum \textit{a posteriori} estimator.   
First, the state $\vect{\psi}$ and observations $\vect{d}$ are assumed to be realizations of their corresponding random variables. 
The confidence is quantified by a probability density function $\mathcal{P}$. 
Second, we assume that our model is Markovian, i.e., our prior on the state follows  
$ 
\mathcal{P}\left(\vect{\psi}(t_k) | \vect{\psi}(t_{k-1}),\vect{F}\right) = \mathcal{P}\left(\vect{\psi}(t_k) | \vect{\psi}\left(t_{1}\right), \dots, \vect{\psi}(t_{k-1}),\vect{F}\right)$. 
%
Third, we assume that the observations $\vect{d}(t_k)$ are conditionally independent given the current state $\vect{\psi}(t_k)$, meaning that the measurement errors are independent and uncorrelated in time. Thus, our \textit{prior} can be updated using the Bayes' rule~\cite{bayes1763essay}
\begin{equation}\label{eq:bayes}
    \mathcal{P}\left(\vect{\psi}(t_k) \mid \vect{d}(t_k),\vect{F}\right) \propto \mathcal{P}\left(\vect{d}(t_k) \mid \vect{\psi}(t_k),\vect{F}\right) \mathcal{P}\left(\vect{\psi}(t_k) \mid \vect{\psi}(t_{k-1}),\vect{F}\right),
\end{equation}
where $\mathcal{P}\left(\vect{d}(t_k) \mid \vect{\psi}(t_k),\vect{F}\right)$ is the likelihood of the observations, i.e., how consistent the predicted state $\vect{\psi}(t_k)$ is with the new observation $\vect{d}(t_k)$; and $\mathcal{P}\left(\vect{\psi}(t_k) \mid \vect{d}(t_k),\vect{F}\right)$ is the resulting \textit{posterior} distribution. 
Fourth, we assume a Gaussian prior (i.e., the forecast) and likelihood functions with  
covariances  
$\matr{C}_{\psi\psi}^f$ and $\matr{C}_{dd}$, such that the posterior remains Gaussian and the analysis state can be computed as 
\begin{align}\label{eq:argmin}
    \vect{\psi}^a = \underset{\vect{\psi}}{\text{arg min}}~{\mathcal{P}\left(\vect{\psi} \mid \vect{d}, \vect{F}\right)} = \underset{\vect{\psi}}{\text{arg min}}~{\mathcal{J}\left(\vect{\psi}\right)}, 
\end{align}
where the cost function to minimize is 
\begin{equation}\label{eq:cost_function_kf}
    \mathcal{J}(\vect{\psi}) =  (\vect{\psi}^f -  \vect{\psi})^\top (\matr{C}_{\psi\psi}^{f})^{-1} (\vect{\psi}^f -  \vect{\psi}) + \left(\vect{d} -  \matr{M}\vect{\psi}\right)^\top \matr{C}_{dd}^{-1} \left(\vect{d} -  \matr{M}\vect{\psi}\right). 
\end{equation} 
For nonlinear $\mathcal{M}$ or $f_2$, the augmented state formulation ensures the measurement operator remains linear in the $\vect{\psi}$ space, preserving the Gaussian update structure.  

\subsection{The ensemble Kalman Filter}\label{sec:enKF}
The ensemble Kalman filter (EnKF) approximates the \textit{prior} and \textit{posterior} distributions via Monte Carlo sampling, i.e., by generating $N_\text{ens}$ instances (ensemble members) of the ROM. 
The EnKF offers a computationally-cheap alternative to the Kalman filter for nonlinear models~\cite{evensen2003ensemble, evensen2009ensemble}. 
In the ensemble framework, each ensemble member represents a different realization of the nonlinear model, thereby capturing nonlinear error growth propagating the full covariance matrix. This makes the EnKF advantageous for high-dimensional systems. 
Unlike other assimilation methods such as the extended Kalman filter or the 4D-Var~\cite{courtier1994strategy}, the EnKF effectively handles nonlinearities without computing tangent linear operators or linearizing the forecast model.

%
The ensemble statistics approximate the expectation and covariance of the model predictions as 
\begin{align}\label{eq:ensemble}
    \mathbb{E}\left(\vect{\psi}\right) &\approx \overline{\vect{\psi}} = \dfrac{1}{N_\text{ens}}\sum_{j=1}^{N_\text{ens}}\vect{\psi}_j,\\
    \matr{C}_{\psi\psi} = \begin{bmatrix}
        \matr{C}_{rr} &  \matr{C}_{r\hat{u}} \\
        \matr{C}_{\hat{u}r} & \matr{C}_{\hat{u}\hat{u}}
    \end{bmatrix} &\approx   \dfrac{1}{N_\text{ens}-1}{ \left(\matr{A}^f - \overline{\vect{\psi}}\vect{1}_{N_\text{ens}}^\top\right)\left(\matr{A}^f - \overline{\vect{\psi}}\vect{1}_{N_\text{ens}}^\top \right)^\top},
\end{align}
where $\matr{A}=[\vect{\psi}_1, \dots,  \vect{\psi}_{N_\text{ens}}]\in\mathbb{R}^{N_\psi\times N_\text{ens}}$ denotes the collection of ensemble members and $\vect{1}_{N_\text{ens}}\in\mathbb{R}^{N_\text{ens}}$ is a column vector of ones.  
Additionally, under linear-Gaussian assumptions, the ensemble approximation preserves the first two moments exactly, which ensures consistency with the Kalman filter in the limit of infinite ensemble size.  
At every assimilation step, each ensemble member $\vect{\psi}_j^f$ is updated by minimizing the cost function~\eqref{eq:cost_function_kf}, which has the analytical solution 
\begin{equation}\label{eq:EnKF}
    \vect{\psi}_j^a = \vect{\psi}_j^f + \underbrace{\matr{C}_{\psi\psi}^f \matr{M}^\top\left(\matr{M} \matr{C}_{\psi\psi}^f \matr{M}^\top + \matr{C}_{dd}\right)^{-1}}_{\text{Kalman gain}}\left(\vect{d}_j - \matr{M}\vect{\psi}_j^f\right), \; \text{for} \; j=1, \dots, N_\text{ens}.    
\end{equation}
Each ensemble is updated with a different  $\vect{d}_j \sim \mathcal{N}\left(\vect{d}, \matr{C}_{dd}\right)$ to avoid underestimating the analysis covariance~\cite{anderson2012localization}.  
The EnKF~\eqref{eq:EnKF} can written in matrix form as 
\begin{align}\nonumber
\matr{A}_r^a &= \matr{A}_r^f + \matr{C}_{r\hat{u}} \left(\matr{C}_{dd} + \matr{C}_{\hat{u}\hat{u}}\right)^{-1} \left(\matr{D} - \matr{A}_{\hat{u}}^f \right)
\\
&=\matr{A}_r^f  \left[\mathbb{I}_m + \frac{1}{N_\text{ens}-1}\left(\matr{A}^f_{\hat{u}}\right)^\top \left(\matr{C}_{dd} + \matr{C}_{\hat{u}\hat{u}}\right)^{-1} \left(\matr{D} - \matr{A}^f_{\hat{u}} \right)\right]
\end{align}
where $\matr{A}_r$ and  $\matr{A}_{\hat{u}}$ are the rows of the ensemble $\matr{A}$ containing the reservoir states $\vect{r}_j$ and the observables $\mathcal{M}\left(\vect{\hat{u}}_j\right)$, respectively, 
and  $\matr{D}=[\vect{d}_1, \dots, \vect{d_{N_\text{ens}}}]$. 
This EnKF formulation avoids the storage of $\mathcal{O}(N_\psi^2)$ matrices, reducing the cost to $\mathcal{O}\left(N_\text{ens}\times N_r\right)$. 
For further details on the EnKF and data assimilation, we refer to~\cite{evensen2009data, law2015data, sanz2023inverse}. 


\section{DA-CAE-ESN implementation}\label{sec:implement}

The DA-CAE-ESN framework is algorithmically implemented as follows.
\begin{enumerate}[align=parleft,labelwidth=1.5em,labelsep=0.5em,itemindent=0pt,leftmargin=!]
\item \textbf{CAE-ESN set-up}
\begin{enumerate}
\item \textit{CAE training:} Given training data $\{\vect{u}(t_k)\}_{k=1}^{N_\text{train}}$, train the CAE to 
minimize the MSE loss (Eq.~\eqref{eq:AE}). This yields the low-dimensional latent representation of the training dataset $\vect{y}(t_k) = \mathcal{E}(\vect{u}(t_k)) \in \mathbb{R}^{N_\text{lat}\ll N}$ for $k=1,\dots, N_\text{train}$.
The physical state is approximated as $\vect{\Hat{u}}(t_k)\approx \mathcal{D}(\vect{y}(t_k))\in\mathbb{R}^{N}$. 
\item \textit{ESN training}: Given the training latent trajectories $\{\vect{y}(t_k)\}_{k=1}^{N_\text{train}}$, train the ESN to learn the temporal mapping $\vect{y}(t_k) \to \vect{\hat{y}}(t_{k+1})$. 
\item  \textit{Assimilation-free forecast}: 
    Initialize $\vect{r}(t_0)=\vect{0}$ and perform $N_\text{wash}$ open-loop steps using the encoded snapshots as input. Next, run the ESN in closed-loop for a transient time and decode the latent trajectory to obtain the physical state forecast.
\end{enumerate}

\item \textbf{Initialization}
\begin{enumerate}
\item \textit{Ensemble initialization} $\{\vect{\psi}_j^f\}_{j=1}^{N_\text{ens}}$:  In physics-based models, ensemble members are generated by adding uncertainty to the equation-based model state and parameters using physical information~\citep[e.g.,][]{papadakis2010data}. Here, we construct an ensemble of latent states using perturbed initial conditions or by randomly sampling states from the training set.
\item \textit{Reservoir washout}: Each reservoir is initialized as $\vect{r}_j(t_0)=\vect{0}$. Then, for each ensemble member, we compute the reservoir for $N_{\text{wash}}$ open-loop steps, using the encoded snapshots from the ensemble initialization. The washout phase is not used for forecasting but it drives the reservoir state towards the system's attractor. This mitigates the influence of the zero initialization and stabilises the subsequent forecast.
\end{enumerate}
\item \textbf{Data assimilation cycle} (Repeat until convergence/criteria met)
\begin{enumerate}
    \item \textit{Forecast step}: Integrate each ensemble member through the nonlinear model 
    $\vect{\psi}_j^f(t_{k+1}) = \vect{F}(\vect{\psi}_j^a(t_k))$, for $j=1,\dots,N_\text{ens}$, until observations $\vect{d}$ become available. 
    This operation can be parallelized for computational efficiency.  
    \item \textit{Analysis step}: Sample  $\vect{d}_j\sim\mathcal{N}\left(\vect{d}, \matr{C}_{dd}\right)$ and obtain the analysis ensemble  via the EnKF~\eqref{eq:EnKF}. We select $\matr{C}_{dd} = \operatorname{diag}(\sigma \vect{u}_{\text{max}})$, where $\vect{u}_{\text{max}}$ is the element-wise maximum of $\{\vect{u}(t_k)\}_{k=1}^{N_\text{train}}$.
    \item \textit{Inflation} (applied when necessary):
    To mitigate ensemble covariance  underestimation (i.e., covariance collapse) due to e.g.,  finite ensemble size,  we apply additive covariance inflation
    \begin{equation*}
   \vect{\psi}_j^a = \overline{\vect{\psi}}^a +   \rho_I\left(\vect{\psi}_j^a - \overline{\vect{\psi}}^a\right), \quad \text{for } j=1,\dots,N_\text{ens}
        \end{equation*}    
    where $\rho_I \geq 1$ is inflation factor~\cite{evensen2009data}. We use $\rho_I = 1.1$ or $1.05$ in selected cases; most cases proceed without inflation, i.e., $\rho_I=1.0$. 
    \item \textit{Re-initialization:}  The analysis ensemble members $\{\vect{\psi}_j^a\}_{j=1}^{N_\text{ens}}$ are the initial conditions for the next forecast step.  
\end{enumerate}

\item {\textbf{Decoding and post-processing}: 
At any time step $t_k$ of the assimilation cycle, the physical flow field can be decoded from the latent state forecast. 
\begin{enumerate}
    \item \textit{Mean reservoir state}: The best estimate in the EnKF is the ensemble mean. Thus, we first compute the mean of reservoir states $\overline{\vect{r}}(t_k) = \frac{1}{N_\text{ens}}\sum_{j=1}^{N_\text{ens}}\vect{r}_j(t_k)$. 
    \item \textit{Decoding}: To find the physical state prediction from $\overline{\vect{r}}$, we compute the latent state with the readout~\eqref{eq:esn2} as $\vect{\hat{y}}(t_k) = \matr{W}_{\text{out}}\overline{\vect{r}}(t_k)$, and then decode $\vect{\hat{u}}(t_k) = \mathcal{D}\left(\vect{\hat{y}}(t_k)\right)$. 
    \item \textit{Visualization and analysis}: Generate visualizations and statistical metrics to assess the quality of the forecast, which we evaluate with the mean absolute error (MAE) at each time step $t_k$, defined as
    \begin{align}
    \text{MAE}_k(\vect{u}, \vect{\Hat{u}}) = \frac{1}{N}\| \vect{u}\left(\vect{x}, t_k\right) - \vect{\hat{u}}\left(\vect{x}, t_k\right)\|_1,
    \end{align}
    where $\|\cdot\|_1$ refers to the $\ell_1$ norm. The MAE over a given time interval is obtained by averaging $\text{MAE}_k$ across the respective time steps.
\end{enumerate}}
\end{enumerate}

In sections \S\ref{sec:ks-results} and \S\ref{sec:results-kolmogorov}, we show the DA-CAE-ESN on the Kuramoto-Shivasinky equation and the 2D Navier-Stokes, respectively. 
Table~\ref{tab:cae-esn-enkf-details} summarizes the parameters employed in the simulations. 

\begin{table}[h]
\centering
\caption{DA-CAE-ESN parameters used for the KS equation and the Kolmogorov flow (2D Navier-Stokes) on quasi-periodic (QP) and chaotic (CH) regimes.}
\begin{tabular}{@{}llccc@{}}
\toprule
\multicolumn{2}{l}{\textbf{Parameter} }
& \begin{tabular}{c}\textbf{KS Equation} \\ (\S\ref{sec:ks-results})\end{tabular}& \begin{tabular}[c]{@{}c@{}}\textbf{Kolmogorov} \\ QP (\S\ref{sec:results-kol-qp})\end{tabular} & \begin{tabular}[c]{c}\textbf{Kolmogorov}\\ CH (\S\ref{sec:results-kol-ch})\end{tabular} \\\midrule
$N_\text{train}$ & Train snapshots &  50,000 & 50,000& 50,000\\
- & Validation snapshots &20,000&20,000&20,000\\
- & Test snapshots & 50,000 & 50,000 & 150,000 \\

$N_\text{wash}$ & Washout time steps & 100 & 100 & 100\\
$N_r$ & Reservoir size & 10,000 & 10,000 &  20,000 \\
$N_\text{lat}$ & Latent dimension & 24 & 8 &  48 \\\midrule
$N_\text{obs}$ & Number of sensors & 25 & 64  &  144\\
-  & Full state dimension & 128 & 2304 (48x48) &   2304 (48x48)  \\
$N_\text{ens}$ & Ensemble size & 100 &  50& 50 \\
$\sigma$ & Noise level & 0.1 & 0.1 & 0.1 \\
$\Delta t_\text{obs}$ & Observation interval & $1 \tau_\lambda$ &  10~TUs & 1~TU\\\midrule
$Re$ & Reynolds number & - & 30 & 34 \\
$\alpha$ & Drag amplitude & - & 0.0 & 0.1\\
$\Delta t$ & Snapshot time step& 0.25 & 0.1 & 0.1\\
\bottomrule
\end{tabular}
\label{tab:cae-esn-enkf-details}
\end{table}

\section{Kuramoto-Sivashinsky equation}\label{sec:ks-results}
The Kuramoto-Sivashinsky (KS) equation is a one-dimensional partial differential equation (PDE) that describes diffusion-induced spatio-temporal chaos~\cite{Hyman_1986_KuramotoSivashinsky}. For a range of paratmeters, the dynamics of the KS equation have persistent, chaotic behaviour over space and time, providing a canonical example of spatiotemporal chaos relevant to a wide range of fluid flow phenomena~\cite{kalogirou2015depth},
such as flame front propagation~\cite{kuramoto1978diffusion},
instabilities of thin liquid films flowing down an inclined plane~\cite{nepomnyashchii1974stability},
or diffusion-induced chaos in reactive systems 
~\cite{kuramoto1976persistent}. 
Writing $\vect{u} =\vect{u}\left(t,\vect{x}\right)$, the KS equation is given by
\begin{equation}\label{eq:ks_equation}
\vect{u}_t +\vect{u}_{xx} + \vect{u}_{xxxx} +\vect{u}\vect{u}_{x} = 0, 
\end{equation} 
with periodic boundary conditions $\vect{u} \left(t, 0\right) =\vect{u} \left(t, L\right)$ on the domain $[0, L]$. The spatial domain is discretized as $ \vect{x} = [x_1; \dots; x_N]$. The KS system is chaotic for $L \geq 22$~\cite{ Hyman_1986_KuramotoSivashinsky,papageorgiou1991route}. Here, we consider the chaotic KS equation with  $L = 20\pi$, for which the system has a Lyapunov exponent $\lambda \approx 0.08$~\citep[e.g.,][]{ozalp2025stability}. The Lyapunov exponent $\lambda$ defines a characteristic timescale of predictability~\cite{boffetta1998extension}, $\tau_{\lambda} = \lambda^{-1} \approx 12.5$, which quantifies a scale over which two nearby trajectories diverge. We discretize the system with 128 degrees of freedom and solve it with a fourth-order spectral solver for stiff PDEs~\cite{Kassam_2005_fourth_order}.

We train the CAE-ESN with $N_\text{lat} = 24$ and $N_r = 10,000$, following the configuration of \cite{ozalp2025stability}.  
Next, we initialize an ensemble with $N_\text{ens} = 100$ by selecting random washout windows from the training data. Each washout window is first encoded into the latent space and used for the warm-up phase of the reservoir of the ESN ensemble. From these random initializations on the attractor, the ESN ensemble is evolved in closed-loop within the latent space for one $\tau_{\lambda}$. 
Then, we start the data assimilation loop.


\subsection{Washout-free CAE-ESN}
\begin{figure}[!htb]
    \centering
    \includegraphics[width=0.95\linewidth]{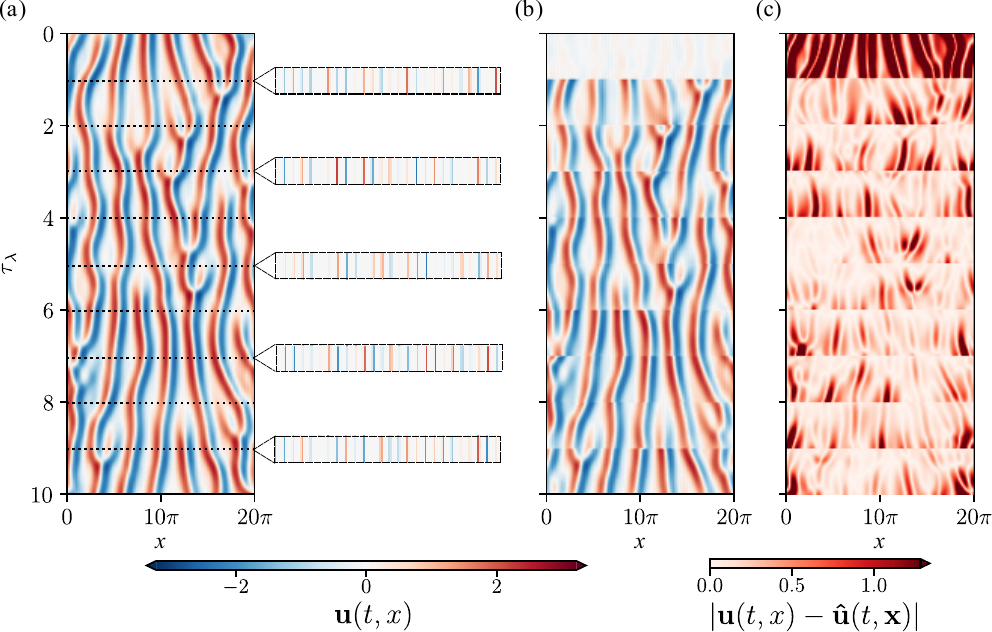}
    \caption{Kuramoto-Sivashinsky dynamics. Comparison between (a) the test data with (b) the reconstruction from the DA-CAE-ESN, and (c) the error reconstruction. The CAE-ESN is initialized randomly, and the assimilation begins at $\tau_\lambda=1$. The $N_\text{obs}=25$ measurements are indicated with the dashed line in (a), and are extended vertically for visualization. 
    }
    \label{fig:ks_cae_esn_enkf_measurements}
\end{figure}

The results are for $N_\text{obs}=25$ with observations sampled every $\Delta t_\text{obs}=\tau_{\lambda}$. Fig.~\ref{fig:ks_cae_esn_enkf_measurements}a shows the measurement locations and Fig.~\ref{fig:ks_cae_esn_enkf_measurements}b shows the mean assimilated forecast obtained using the CAE-ESN. 
Each ensemble member is initialized from a randomly selected washout window in the training data, which leads to an initially zero-mean latent state. As a result, the decoded output remains near zero until the first assimilation at $t = 1 \tau_{\lambda}$. As measurements are assimilated, the error decreases at each observation point (Fig.~\ref{fig:ks_cae_esn_enkf_measurements}c). This allows the assimilated CAE-ESN prediction to converge towards the test data. The DA-CAE-ESN never has access to the full state test data; instead, it is randomly initialized and only receives partial measurements at each $\tau_\lambda$ interval.

\begin{figure}[!htb]
    \centering
    \includegraphics[width=0.95\linewidth]{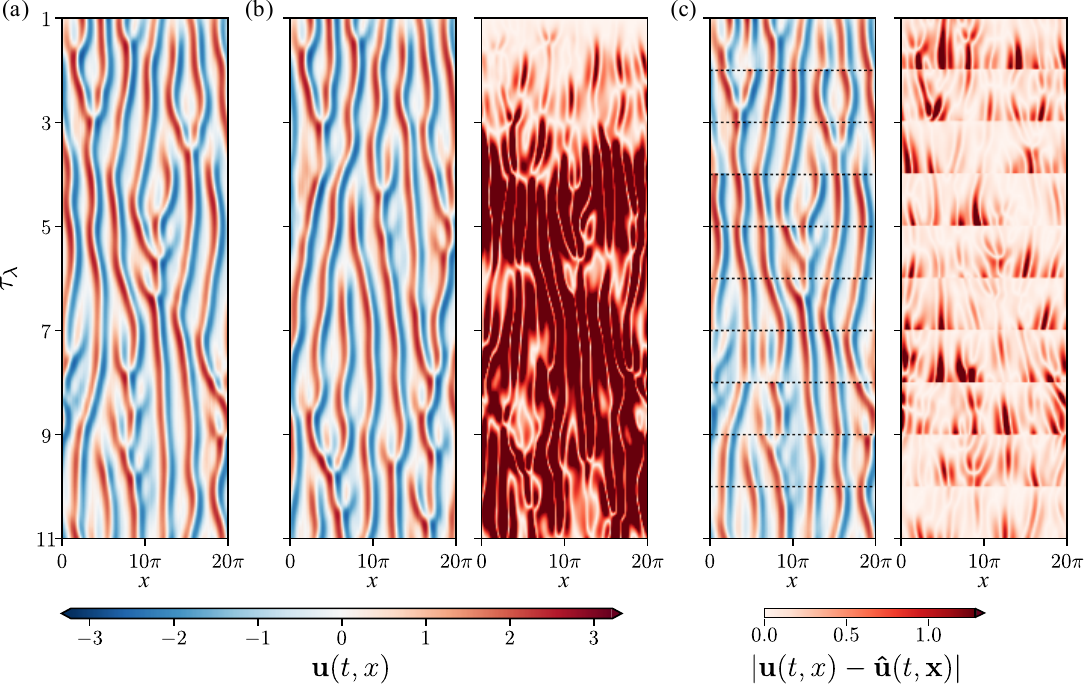}
    \caption{Kuramoto-Sivashinsky dynamics. Comparison of the test data set (a) with the predictions and their corresponding mean average error from (b) the CAE-ESN and the (c) assimilated DA-CAE-ESN. The dotted line in (c) indicates the $25$ measurements.}
    \label{fig:ks_caeesn_vs_enkf_comp}
\end{figure}

\begin{figure}[!htb]
     \includegraphics[width=\textwidth]{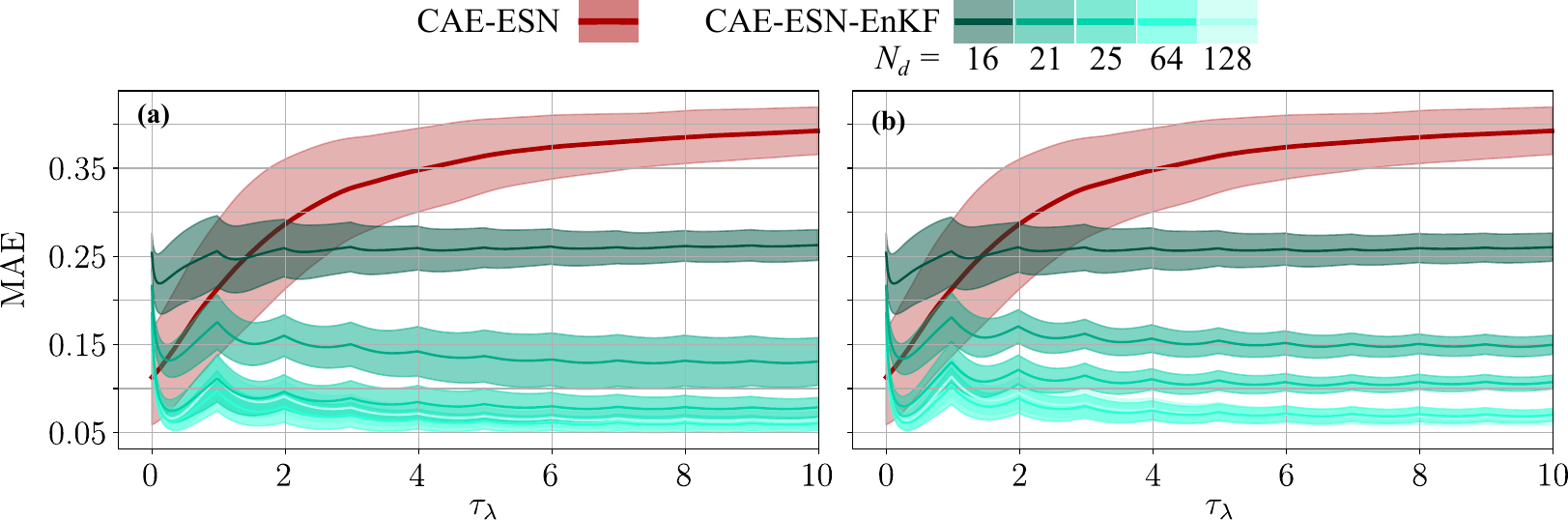}
      
        \caption{Kuramoto-Sivashinsky equation. Comparison between the time evolution of the MSE from the CAE-ESN (red) and the DA-CAE-ESN with different numbers of observations (colormap) with measurement noise  (a) $\sigma= 0.1$, and (b) $\sigma= 0.2$. 
        }
       \label{fig:ks_caeesn_mae_time_samples}
\end{figure}

In Figure~\ref{fig:ks_caeesn_vs_enkf_comp}b, we apply the CAE-ESN to the corresponding test washout window and predict for $10\tau_\lambda$ in closed-loop, without assimilated measurements. For the first two $\tau_\lambda$, the CAE-ESN forecast closely follows the reference test data, but eventually diverges due to the chaotic dynamics of the system, consistent with the findings of~\cite{linot2022data, ozalp2025stability}.
In comparison, the DA-CAE-ESN forecast after the first measurement is shown in Figure~\ref{fig:ks_caeesn_vs_enkf_comp}c. Although the initial error is larger, owing to the limited partial measurements available to the model, the forecast trajectory closely follows the reference test data trajectory. This behaviour is further seen when examining the MAE. In Fig.~\ref{fig:ks_caeesn_mae_time_samples}, we compare the MAE over time between the CAE-ESN and the DA-CAE-ESN forecast. Initially, the MAE of the CAE-ESN is smaller but increases as the system diverges.
Therefore, the augmented state-space assimilation framework  (c.f., \S\ref{sec:da-problem-statement}) allows us to successfully extract information and forecast from partial observations.

\subsection{Parametric study}

We investigate the accuracy with respect to three parameters: 
(i) the number of measurements, 
(ii) the sampling frequency of the measurements, and 
(iii) the noise level. 
The tested parameter values are summarized in Table~\ref{tab:sensitivity_parameters}. To robustly assess the sensitivity of the prediction to the parameters while accounting for the system’s chaotic nature, we independently evaluate the DA-CAE-ESN on $100$ randomly selected intervals of the test data. For $N_\text{obs} = 128$ and $N_\text{obs} = 64$, covariance inflation was applied to prevent the ensemble from collapsing onto the reference measurements (see \S\ref{sec:implement}).
\begin{table}
\centering
\begin{tabular}{l l}
\hline
\textbf{Parameter} & \textbf{Values} \\ \hline
Number of sensors $N_\text{obs}$ & $16, 21, 25, 32, 64, 128$ \\
Sampling frequency $\Delta t_\text{obs} [\tau_\lambda]$& $0.25, 0.5, 0.75, 1, 1.5, 2, 3 $ \\
Noise level & $0.1, 0.2$ \\ \hline
\end{tabular}
\caption{Kuramoto-Sivashinsky equation: Parameters analysed for the DA-CAE-ESN forecast.}
\label{tab:sensitivity_parameters}
\end{table}

First, we explore how many measurements are required for a successful time-accurate forecast. For this, we fix the sampling frequency to $1 \tau_\lambda$, test two noise levels, and vary the number of available observations. In Fig.~\ref{fig:ks_caeesn_vs_enkf_comp}, we show the evolution of the MAE over $10 \tau_\lambda$, averaged over $100$ intervals, with the shaded area indicating the standard deviation. When the full state is assimilated ($N_\text{obs}=128$), the initial MAE is $\approx 0.18$, which is larger than the MAE of the traditional CAE-ESN (MAE$\approx 0.1$), which has been initialized with a full state test window. This difference arises because the DA-CAE-ESN is initialized on random intervals and must rely on measurements to gradually synchronize with the solution. Therefore, a smaller number of measurements results in a larger initial error. Each new assimilation of measurements (every $1 \tau_{\lambda}$) decreases the error until the MAE converges over time. 
When using $25$, or more, measurements, the error stabilizes at a similar level, which shows that assimilating this number of measurements, or more, is a suitable choice. In contrast, the CAE-ESN forecast, as the forecast from other architectures such as LSTMs~\cite{vlachas2022multiscale, ozalp2023reconstruction}, diverges after a few $\tau_{\lambda}$ due to chaos.
\begin{figure}[!htb]
    \centering
    \includegraphics[width=\linewidth]{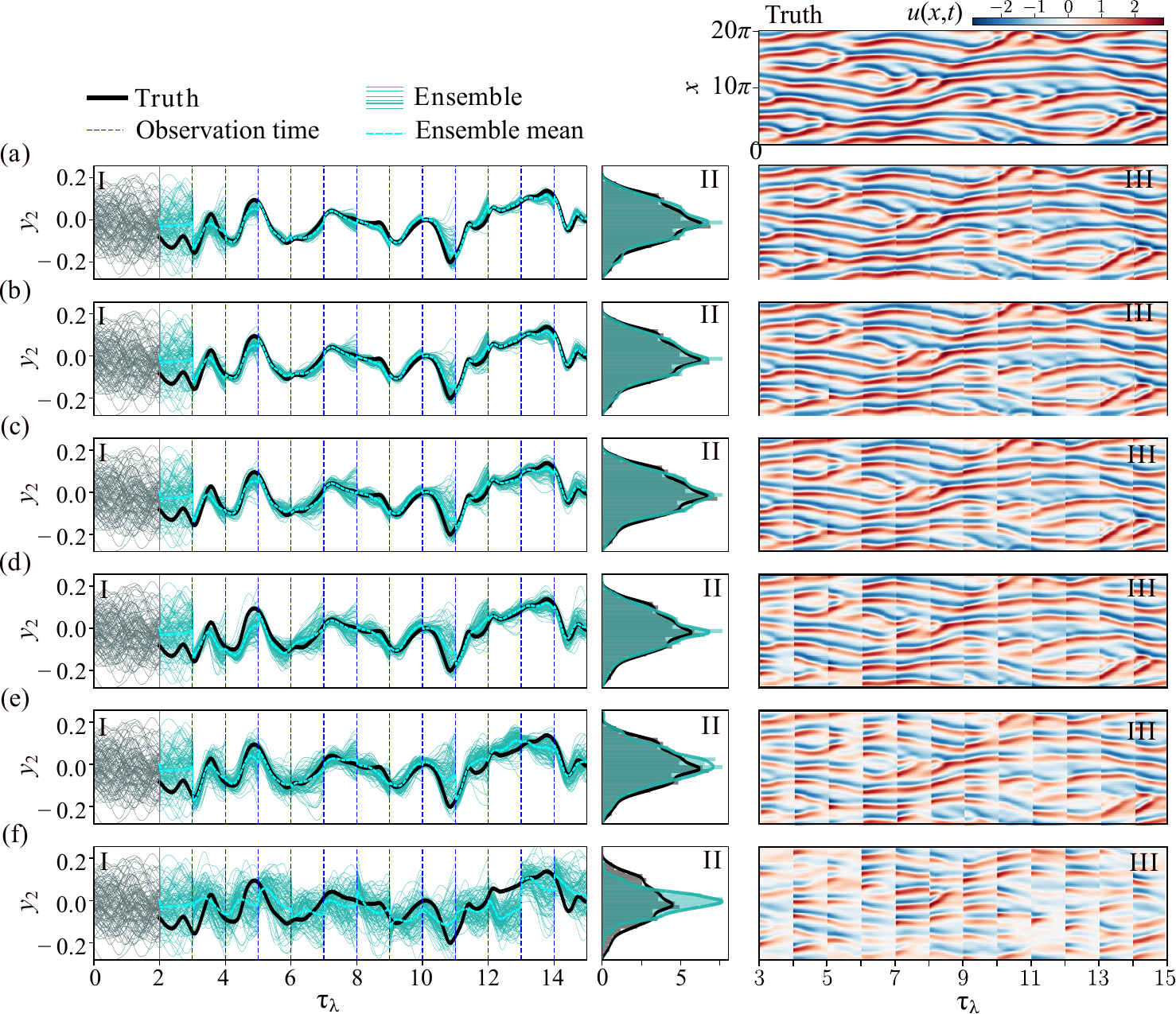}
 \caption{Kuramoto-Sivashinsky equation. Performance of DA-CAE-ESN forecasts for varying number of sensors: $N_\text{obs}=$ (a) 128, (b) 64, (c) 32, (d) 25, (e) 21, (f) 16 at the level of $\sigma=0.2$. 
 (I) Time evolution of the second latent variable over $15 \tau_{\lambda}$. 
 The initial $2 \tau_{\lambda}$ correspond to the random encoded washout. After initializing the reservoir, the ESN predicts in closed-loop for $1 \tau_{\lambda}$ before the first measurements are assimilated.
 (II) Probability density function (PDF) of the truth and the ensemble mean over a long horizon of $100 \tau_{\lambda}$. 
 (III) Decoded DA-CAE-ESN forecasts in the physical domain.
 }\label{fig:ks_caeesnenkf_varsamples_lat02}
\end{figure}

\begin{figure}[!htb]
\includegraphics[width=\textwidth]{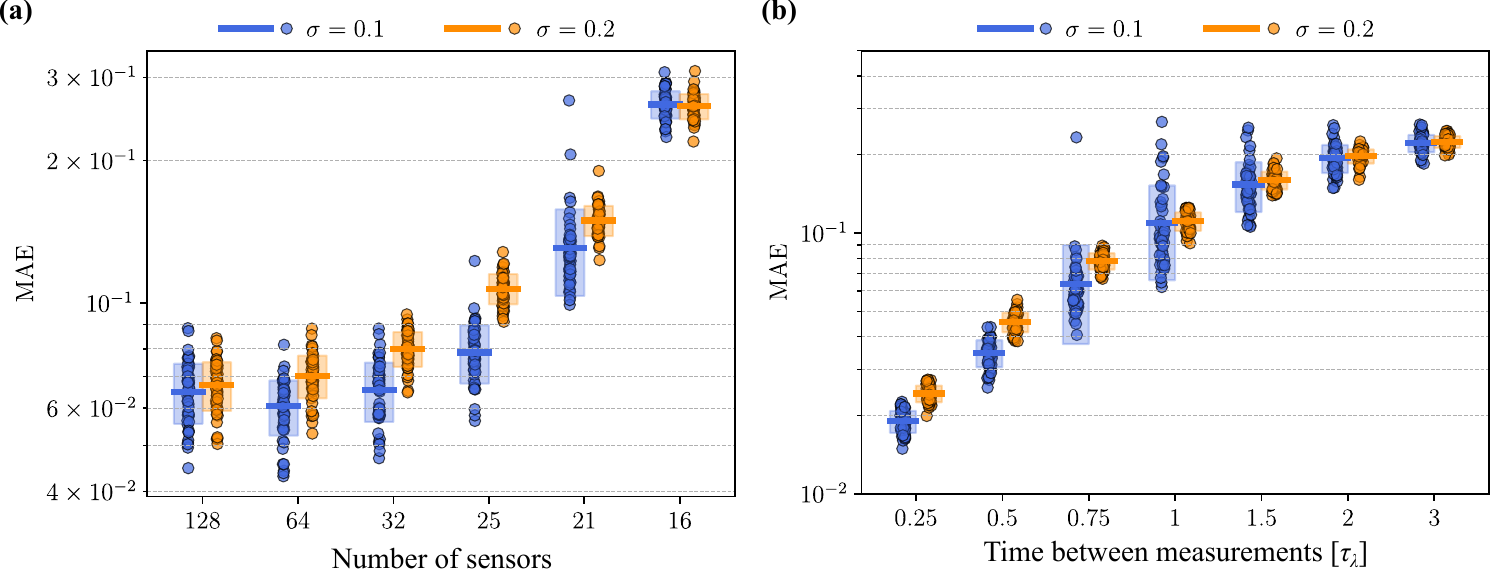}
\label{fig:ks_caeesnenkf_mae_samples0102}
        \caption{Kuramoto-Sivashinsky equation. 
        Reconstruction error at the end of the assimilation when varying 
        (a) the number of measurements at a fixed $\Delta t_\text{obs}=\tau_\lambda$, and 
        (b) the sampling frequency at a fixed $N_\text{obs} = 25$.  
        Results shown for measurement noise $\sigma \in \{0.1, 0.2\}$. }
        \label{fig:ks_caeesnenkf_rmse_samples}
\end{figure}

To further analyse the impact of measurement noise, we analyse the MAE over $10 \tau_\lambda$ whilst varying levels of measurement noise ($\sigma \in \{0.1, 0.2\}$, see \S\ref{sec:implement}). Fig.~\ref{fig:ks_caeesnenkf_rmse_samples}a shows the MAE of the $100$ intervals, together with the mean and standard deviation. With larger measurement noise, we obtain a larger MAE. Additionally, the error increases as the number of observations decreases.

In Fig.~\ref{fig:ks_caeesnenkf_rmse_samples}b, we examine the effect of varying measurement frequencies whilst keeping the number of observations fixed at $N_\text{obs} = 25$. The results indicate that the smallest error occurs at the highest sampling frequency. As the frequency decreases, the mean error increases, following an exponential growth trend consistent with the chaotic dynamics of the system. When increasing the measurement frequency past $1.5 \tau_\lambda$, the MAE plateaus for different measurement noise levels. Additional figures of the decoded solution are provided in Fig.~\ref{fig:ks_caeesnenkf_varsamples_lat02}. 
Although optimal parameter tuning remains an open problem, the DA-CAE-ESN consistently demonstrates robust performance across a range of noise levels, assimilation frequencies, and numbers of measurements. 
The DA-CAE-ESN addresses key limitations of data-driven ROM forecasting models: without assimilation, these models typically diverge after a short time horizon due to chaos and are highly sensitive to partial observations, often failing to reconstruct long-term dynamics accurately \citep[e.g.,][]{ozalp2023reconstruction}. In contrast, our framework incorporates data assimilation in the latent space, which stabilizes predictions despite noise and partial measurement, while bypassing the need for full state initialisation.

\section{2D Navier-Stokes}\label{sec:results-kolmogorov}
Although the Kuramoto–Sivashinsky equation captures essential features of spatio-temporal chaos, fluid systems have higher dimensional dynamics. 
In this section, we consider the Kolmogorov flow, a model for two-dimensional turbulence. The flow evolves on a doubly periodic domain $[0, L_x] \times [0, L_y]$ and it is forced with an amplitude per unit mass of flow $\chi$. 
The Kolmogorov system is governed by the incompressible Navier–Stokes equations and is non-dimensionalized using the length scale $L_y/2\pi$ and the time scale $\sqrt{L_y / (2\pi \chi)}$. This rescaling transforms the domain to $[0, 2\pi L_x / L_y] \times [0, 2\pi]$~\cite{chandler2013invariant}. 
In this work, we consider the domain $\Omega = [0, 2\pi] \times [0, 2\pi]$ and $\chi = 1$ in the non-dimensionalization, such that the Reynolds number is $Re = 1/\sqrt{\nu}$.
With this, the equations governing the Kolmogorov flow are 
\begin{align}\label{eq:kolmogorov}
\partial_t \vect{u} + \vect{u}\cdot \nabla \vect{u} &=- \nabla p + \nu\nabla^2 \vect{u} +\vect{g}, \\
\nabla \cdot \vect{u} &= 0. 
\end{align} 
Here, $\vect{u}\in\mathbb{R}^2$ is the velocity field, 
$p$ is the pressure, 
$\nu$ is the kinematic viscosity, 
and the external forcing is 
$\vect{g}\left(x\right) = \sin(4 y) \vect{e}_1 - \alpha \vect{u}$, 
where $\vect{e}_i$ denotes the standard basis of vectors in $\mathbb{R}^2$, and 
$\alpha$ is the amplitude of the velocity-dependent drag term~\cite{boffetta2012two, kochkov2021machine}. 
The average energy dissipation $D$ and the local dissipation rate $d(\vect{x}, t)$ are defined as 
\begin{align}\label{eq:global_dissipation}
D(\vect{u}, t) = \frac{1}{(2\pi)^2} \int_{\Omega} d(\vect{u}, \vect{x}, t) d\vect{x}, \;\text{and}\;\;  
d(\vect{u}, \vect{x}, t) = \frac{1}{Re} \| \nabla \vect{u}(\vect{x}, t) \|^2.
\end{align}


We test the DA-CAE-ESN framework on two cases in the Kolmogorov flow: 
(i) a quasiperiodic solution of the system for $Re=30, \alpha=0.0$~\cite{racca2023predicting} (\S\ref{sec:results-kol-qp})  and 
(ii) a 2D chaotic solution of the system at $Re=34, \alpha=0.1$ (\S\ref{sec:results-kol-ch}). 
To solve Eq.~\eqref{eq:kolmogorov}, we use JAX-CFD (\href{https://github.com/google/jax-cfd}{\texttt{google/jax-cfd}}),  which implements a finite volume scheme with a second-order time integration method~\cite{kochkov2021machine}. 
The vorticity, defined as $\vect{\omega} = \nabla \times \vect{u}$, is sampled as snapshots on a $48 \times 48$ grid using a solver time step of $0.01$ and subsequently downsampled to $\Delta t = 0.1$. Table~\ref{tab:cae-esn-enkf-details} includes further details on the simulations' parameters. 
%
%
 Although Lyapunov times can be defined in the Kolmogorov flow, we employ the non-dimensionalised time-units (TUs) because the instantaneous or finite-time Lyapunov exponents have significant fluctuations in chaotic  systems, i.e., up to three times the average Lyapunov exponent~\cite{ vastano1991short, mohan2017scaling}. 
As a result, the mean Lyapunov exponent does not necessarily provide a representative scale of the short-term growth rate of perturbations, particularly in regimes for which the dynamics are intermittent.




\subsection{Quasi-periodic case}\label{sec:results-kol-qp}
\begin{figure}[!htb]
    \centering
    \includegraphics[width=\linewidth]{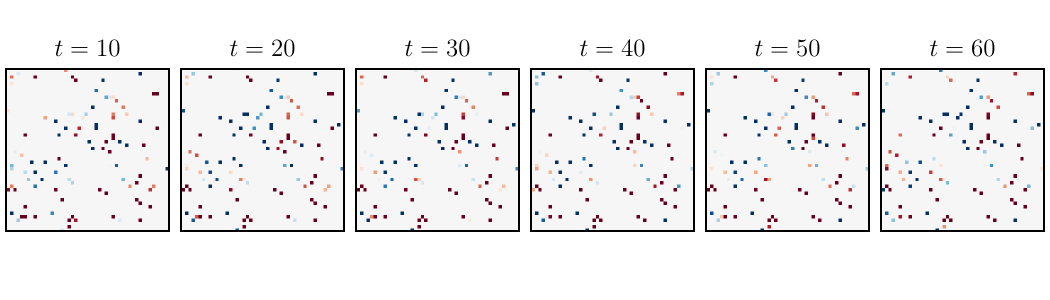}
    \caption{Kolmogorov flow quasi-periodic regime. Sparse observations in the DA-CAE-ESN framework. The vorticity is measured at $50$ spatial locations selected uniformly at random, with additive Gaussian noise of standard deviation $\sigma=0.1$. The colour scale represents the signed vorticity at the measurement points; contrast has been artificially enhanced for visibility.} 
    \label{fig:measurements_10TU_40}
\end{figure}

The quasiperiodic Kolmogorov flow has been analysed by~\cite{platt_kolmogorov_1991, racca2023predicting}, among others. The flow state is quasiperiodic and the dissipation rate is periodic with four neutral Lyapunov exponents~\cite{ozalp2025stability}. 
Here, we train a multi-scale CAE with $N_\text{lat} = 8$ (reducing the number degrees of freedom by a factor of 288) and an ESN with $N_r = 10,000$ to forecast the latent variables. We assimilate vorticity observations every 10~TUs from 50 sensors (corresponding to $2.17\%$ of the full state). The sensors are selected uniformly at random and perturbed with a Gaussian noise of $10\%$ of the standard deviation as shown in Fig.~\ref{fig:measurements_10TU_40}.  

\begin{figure}[!htb]
    \centering
    \includegraphics[width=\linewidth]{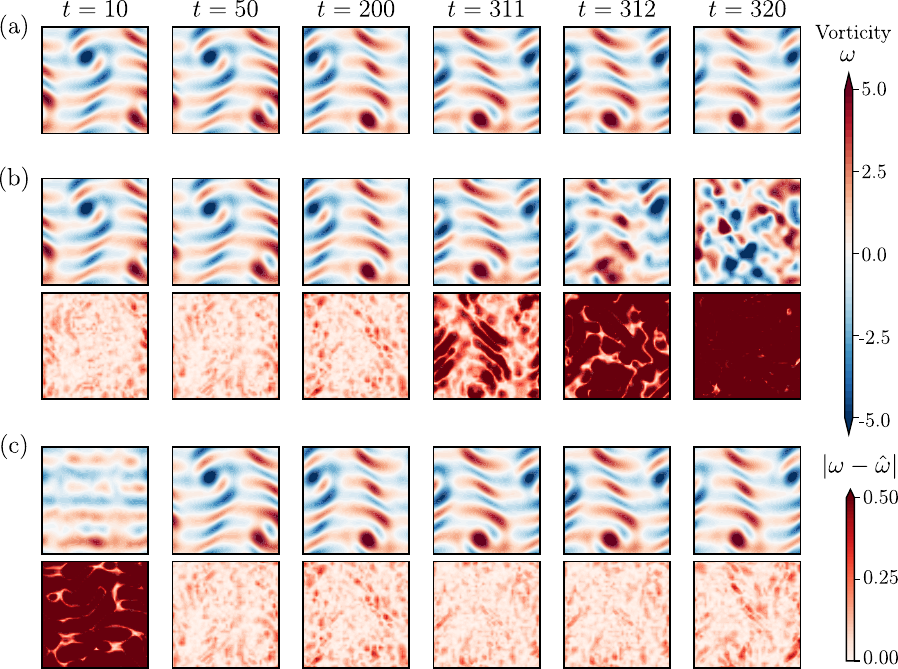}
    \caption{Kolmogorov flow quasi-periodic regime. Comparison of (a) test data with the predictions, and their corresponding mean absolute error over $400$~TU ($4000$ snapshots) for (b) the CAE-ESN and (c) the proposed DA-CAE-ESN.}
    \label{fig:CAEESN_vs_ENKF_10TU_24}
\end{figure}

\begin{figure}[!htb]
    \centering
    \includegraphics[width=0.95\linewidth]{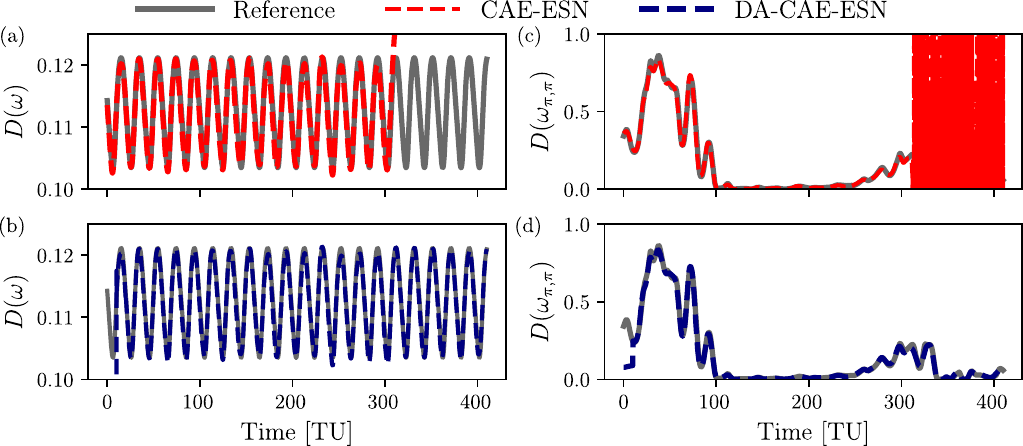}
\caption{Kolmogorov flow quasi-periodic regime. Comparison of dissipation for the reference test data with (a,b) the CAE-ESN, and  (c,d) the DA-CAE-ESN. The dissipation is computed (a, c) globally and (b, d) locally at $\left(\pi, \pi\right)$ over $400$~TUs. Without assimilation, the CAE-ESN forecast becomes unstable, resulting in an exponential divergence around $312$~TUs. }
    \label{fig:CAEESN_vs_ENKF_dissipation_24}
\end{figure}

Figure~\ref{fig:CAEESN_vs_ENKF_10TU_24} compares the evolution of the true vorticity field (Fig.~\ref{fig:CAEESN_vs_ENKF_10TU_24}a) with the predicted fields from the CAE-ESN initialized with the true washout data  (Fig.~\ref{fig:CAEESN_vs_ENKF_10TU_24}b) and the DA-CAE-ESN ensemble mean (Fig.~\ref{fig:CAEESN_vs_ENKF_10TU_24}c).  
Similarly to the KS case, the large error before the assimilation begins ($t=10$~TUs) is due to the initialization of the ensemble on randomly selected washout windows of the training set (see  \S\ref{sec:implement}).  
Once the assimilation begins, the forecast approaches the true state. 
The autonomous CAE-ESN is typically stable for the quasiperiodic regime. However, depending on the initial washout, it may accumulate errors that lead to instability. Specifically, in Fig.~\ref{fig:CAEESN_vs_ENKF_10TU_24}b, we show how the autonomous prediction of the CAE-ESN is accurate in short-term predictions but  at $t \approx 311$~TUs, the accumulated errors cause the vortex structures to break down, which leads to divergence from the true dynamics. 
In contrast, the DA-CAE-ESN prediction in Fig.~\ref{fig:CAEESN_vs_ENKF_10TU_24}c follows the reference trajectory, which demonstrates the ability of the proposed framework to stabilize the CAE-ESN, even without full-state initialization and using noisy sensors data on $2.17\%$ of the domain. 
This is also evidenced by the predicted dissipation rates in Fig.~\ref{fig:CAEESN_vs_ENKF_dissipation_24}. The global dissipation rate from the CAE-ESN has an unbounded growth beyond approximately $310$~TUs, diverging significantly from the periodic dissipation observed in the ground truth. In contrast, when assimilating partial, noisy measurements, both global and local dissipation quantities remain bounded and closely follow the ground truth throughout the entire forecast. 

\subsection{Chaotic Kolmogorov Flow}\label{sec:results-kol-ch}
The Kolmogorov flow at $Re=34$ and $\alpha=0.1$ has a spatio-temporally chaotic behaviour~\citep[e.g.,][]{kochkov2021machine, ozalp2025stability}. In this section, we investigate the effect of the number of sensors and their location on the DA-CAE-ESN forecast accuracy. \\

\begin{figure}[!htb]
    \centering
    \includegraphics[width=\linewidth]{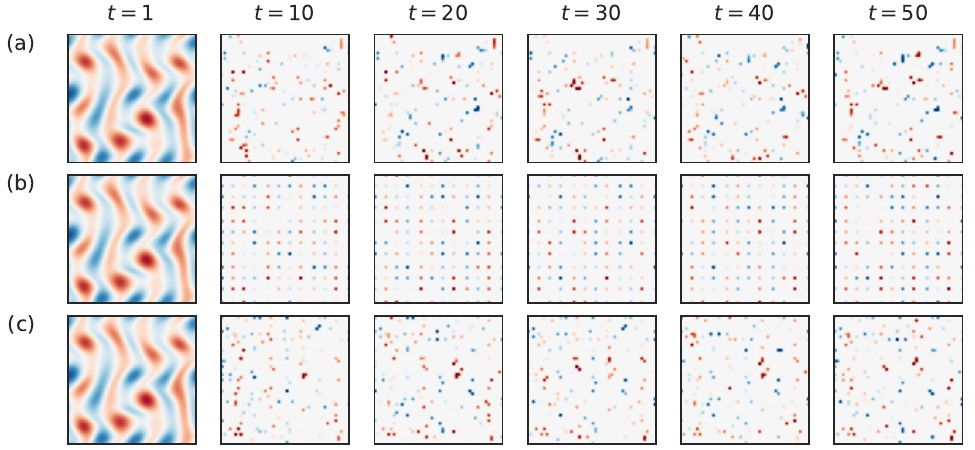}
    \caption{Kolmogorov flow chaotic regime. 
     Observations in the DA-CAE-ESN framework. 
     After an initial input snapshot, the vorticity is measured at $144$ points, which are placed
    (a) randomly, 
    (b) on a $12\times 12$ grid,
    (c) at the maxima and minima of the first $72$ POD modes' energy.  
}
\label{fig:measurements_1TU_34}
\end{figure}

First, although the goal is not optimising sensor placements, we compare three sensor arrangements: 
(i) uniformly random sampling (Fig.~\ref{fig:measurements_1TU_34}a), 
(ii) equidistant placement on a regular grid (Fig.~\ref{fig:measurements_1TU_34}b), 
and 
(iii) placement according to POD (Fig.~\ref{fig:measurements_1TU_34}c). 
In the third strategy, sensors are placed at spatial locations corresponding to the maxima and minima of the leading POD modes, i.e., the regions where the system has the greatest variance. For example, in the case shown in Fig.~\ref{fig:measurements_1TU_34} with $N_{\text{obs}} = 144$ measurements, we take the first $N_{\text{obs}}/2 = 72$ POD modes and select the spatial locations of their maxima and minima. If overlaps occur, additional modes are considered until $N_{\text{obs}}$ sensors are obtained.
 This method is motivated by the goal of maximizing the observability of the system: observations are informative when taken in regions with the largest variance in the dominant modes of the flow~\cite{cohen2003sensor, willcox2006unsteady}. 

\begin{figure}[!htb]
    \centering
    \includegraphics[width=0.85\linewidth]{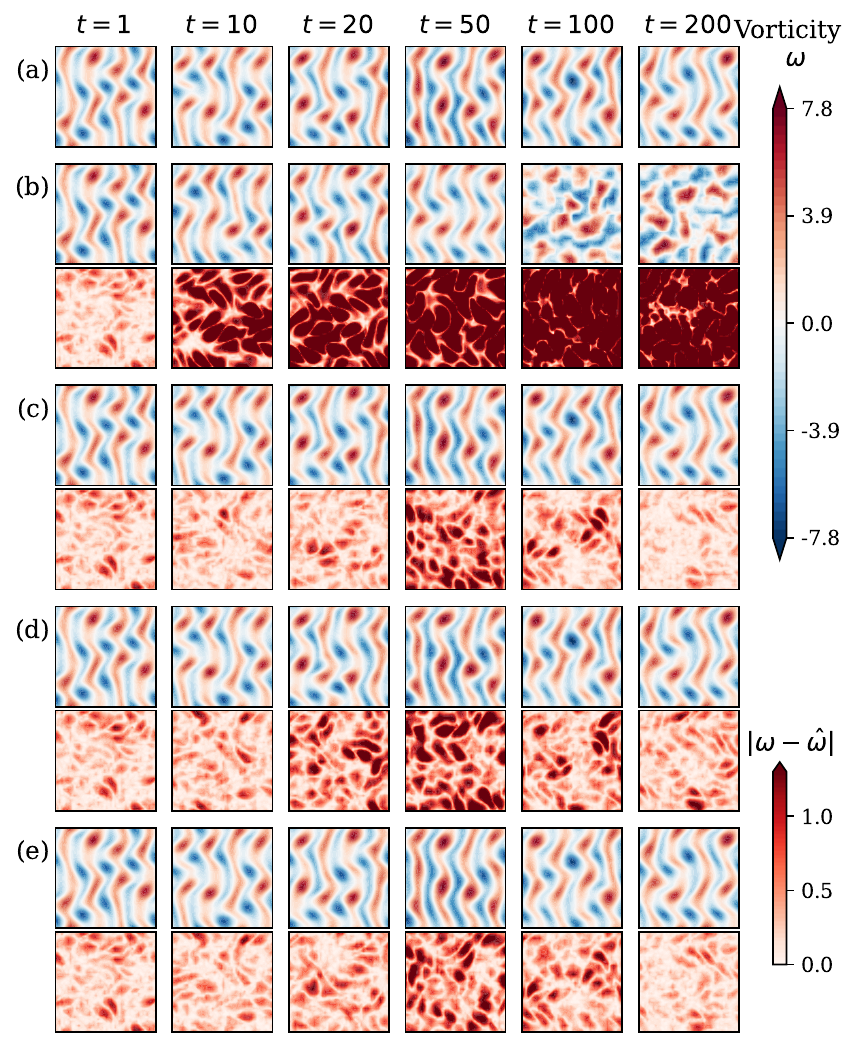}
    \caption{Kolmogorov flow chaotic regime. Comparison of 
    (a) the test data with the predictions and the mean absolute error for 
    (b) the CAE-ESN,  
    the proposed DA-CAE-ESN using sensors placed
    {(c) randomly, 
    (d) on a $12\times 12$ grid,
    (e) at the maxima and minima of the first $72$ POD modes' energy.}  
    }
    \label{fig:CAEESN_vs_ENKF_1TU_34}
\end{figure}

The DA-CAE-ESN prediction under the three sampling scenarios are compared against the assimilation-free CAE-ESN in Fig.~\ref{fig:CAEESN_vs_ENKF_1TU_34}. 
Figure~\ref{fig:CAEESN_vs_ENKF_1TU_34}b shows the autonomous vorticity prediction of the CAE-ESN over 100~TUs on unseen test data. For the initial 10~TUs, the CAE-ESN forecast remains close to the ground truth but then diverges due to chaos. After 50~TUs, the CAE-ESN becomes numerically unstable, which leads to unphysical vorticity predictions.

Because of the increased dynamical complexity and uncertainty, we modify the simulation parameters with respect to the quasi-periodic regime. 
First, we train a CAE-ESN with $N_\text{lat}=48$, compressing the data by a factor of 48, and $N_r = 20,000$~\cite{ racca2021robust, ozalp2025stability}. 
Second, we provide a single full-state noisy snapshot to initialize the reservoir, which is fed $N_\text{wash}$ times to the ESN~\cite{ozan2025data}. This was required because of the sensitivity of the chaotic Kolmogorov flow to perturbations. 
Third, we use 144 sensors, corresponding to $\sim 6\%$ of the full vorticity field, which are assimilated every $1$~TU. 
(Further hyperparameter tuning, e.g., varying ensemble size, noise level, or reservoir size, could improve the results that follow, but the conclusions of the paper remain unchanged.)

\begin{figure}[!htb]
    \centering
    \includegraphics[width=\linewidth]{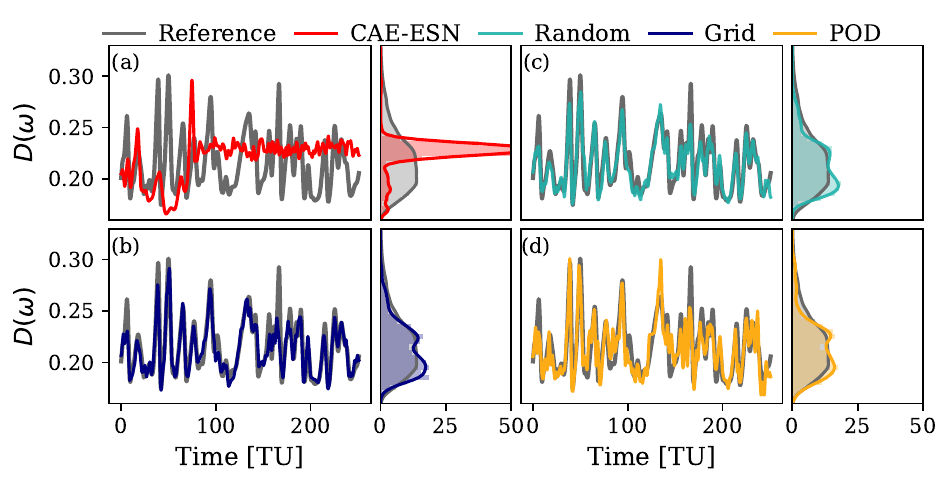} 
    \caption{Kolmogorov flow chaotic regime. Comparison of the global dissipation over $250$~TUs and its distribution for the reference test data, (a) the CAE-ESN and the DA-CAE-ESN using sensors placed
    (b) randomly, 
    (c) on a $12\times 12$ grid,
    (d) on POD modes' energy. 
    } 
    \label{fig:CAEESN_vs_ENKF_dissipation_34}
\end{figure}

In contrast, the DA-CAE-ESN forecasts the structure of the ground truth throughout the prediction horizon (Figs.~\ref{fig:CAEESN_vs_ENKF_1TU_34}c-e). 
The comparable MAE across different sampling strategies indicates that the primary source of error arises from the dimensionality reduction imposed by the autoencoder. The sampling methods yield similar, but not identical forecasts, and all placements result in stable long-term assimilation. 
Although optimal sensor placement and localization techniques are crucial for accurate state reconstruction and traditional data assimilation~\cite{alonso2004optimal,  king2015observability,manohar2018data}, we find that sensor placement has a reduced impact on latent-state assimilation. 
This is because the proposed DA-CAE-ESN assimilates the latent variables, which are a nonlinear combination of the full state and effectively performs a localization of the ensemble covariance. 
These results can also be seen in terms of the energy dissipation of the system in Fig.~\ref{fig:CAEESN_vs_ENKF_dissipation_34}.  
The autonomous CAE-ESN (Fig.~\ref{fig:CAEESN_vs_ENKF_dissipation_34}a) prediction diverges from the truth until it eventually becomes unphysical around $80$~TUs.
In contrast, the energy dissipation resulting from the proposed DA-CAE-ESN's prediction (Figs.~\ref{fig:CAEESN_vs_ENKF_dissipation_34}b-d) is well captured. As observed by \cite{page2021revealing}, who compared dissipation statistics across different latent dimensions, high-dissipation events tend to be under-represented. This discrepancy is due to the smoothing effect of the MSE loss in the autoencoder, which filters out extreme values in the dimensionality reduction to the latent space. 

\begin{figure}[!htb]
    \centering
    \includegraphics[width=.9\linewidth]{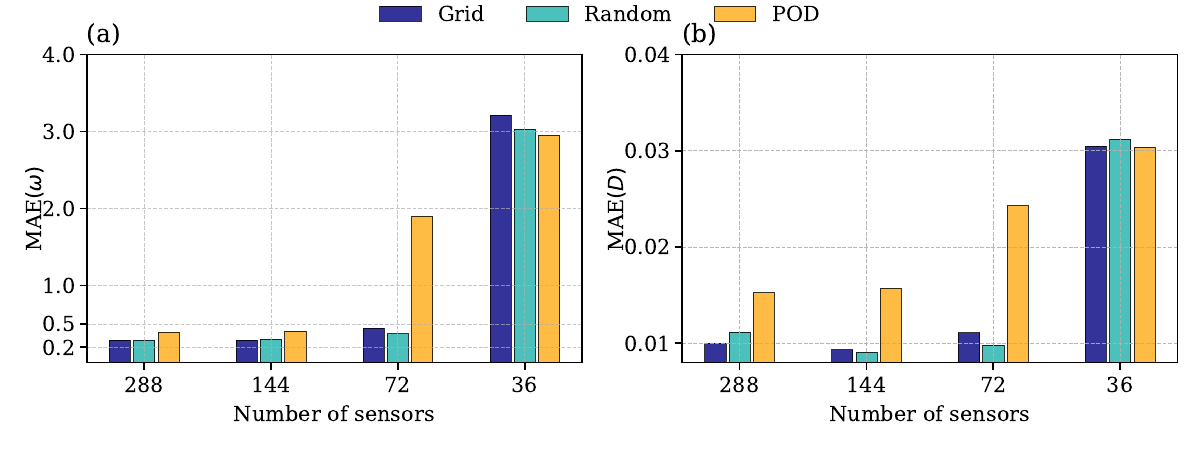}
    \caption{Kolmogorov flow chaotic regime. Effect of the number of sensors and their location on the DA-CAE-ESN prediction. Average mean average error over 250~TUs of the (a) vorticity and (b) the dissipation.}
    \label{fig:kolmogorov_mae_vort_dissip}
\end{figure}

Next, we analyse the effect of the number of observations $N_\text{obs}$ on the DA-CAE-ESN. We compare the performance of the DA-CAE-ESN with a sensor coverage between  1.6\% and 12.5\% of the full state under the different sampling strategies. The details of the sensors' locations are summarised in Table~\ref{tab:kolmogorov_sensitivity_parameters}. 
%
%
Figure~\ref{fig:kolmogorov_mae_vort_dissip} shows the average MAE of the vorticity field prediction (Fig.~\ref{fig:kolmogorov_mae_vort_dissip}a) and the energy dissipation (Fig.~\ref{fig:kolmogorov_mae_vort_dissip}b)  over $250$~TUs for varying numbers of sensors. The error magnitudes are comparable across all placement strategies, except in the POD case at $72$ measurements. Further investigation shows that for this case, the forecast diverges after approximately $200$~TUs. 
This shows that the proposed DA-CAE-ESN yields accurate and numerically stable predictions of the spatio-temporally chaotic systems from sparse and noisy measurements, requiring only $N_\text{obs}>72$, i.e., 3.1\% of the full state. 

Forecast divergence and inherent instability are typical in ROM models without assimilation, where networks like the CAE-ESN tend to blow up rapidly due to chaotic dynamics. In contrast, the DA-CAE-ESN is robust on this 2D chaotic case: across various measurement strategies and different numbers of observations, it successfully stabilizes long-term autonomous forecasts (as seen in Fig.~\ref{fig:CAEESN_vs_ENKF_dissipation_34}).

\begin{table}
\centering
\caption{Kolmogorov flow: Parameters for the DA-CAE-ESN forecast sensitivity analysis}
\label{tab:kolmogorov_sensitivity_parameters}
\begin{tabular}{lcccc}
\toprule
\textbf{Parameter}         & \multicolumn{4}{l}{\textbf{Values}}  \\
\midrule
Number of sensors $N_\text{obs}$   & 288   & 144   & 72    & 36    \\
Measurement percentage (\%) & 12.5  & 6.25  & 3.125 & 1.6   \\
Grid placement (method (ii))  & 16$\times$18   & 12$\times$12   & 8$\times$9     & 6$\times$6     \\
\cmidrule(lr){1-5}
Placement methods & \multicolumn{4}{c}{Random, equidistant grid, POD} \\
\bottomrule
\end{tabular}
\end{table}

\section{Conclusion}\label{sec:conclusion}

Real-time forecasting of high-dimensional chaotic systems typically requires reduced-order models (ROMs). A strategy to design ROMs is to employ autoencoders, which nonlinearly compress the state into a lower-dimensional latent space. Autoencoder-based ROMs have proved successful in a large spectrum of applications; however, they have two limitations: (i) they may become numerically unstable in chaotic regimes; and (ii) they do not naturally handle sparse and noisy measurements in real-time (i.e., they are offline methods). 
To address this, we develop the data-assimilated convolutional autoencoder echo state network (DA-CAE-ESN, with psuedoalgorithm in Sec.~\ref{sec:implement}). This method consists of a three-step strategy:  (i) a convolutional autoencoder (CAE) to compress the high-dimensional state into a latent space;  (ii) an echo state network (ESN) to propagate the dynamics within the latent space;  and (iii) an augmented state-space sequential data assimilation framework (specifically, the ensemble Kalman Filter) to update the ESN forecast by assimilating sparse and noisy measurements into the latent space. The DA-CAE-ESN provides accurate short- and long-term forecasts and reconstructs the full system state without full-state information during assimilation. 
We test the DA-CAE-ESN framework on two spatio-temporally chaotic systems to compare accuracy, robustness and stability against an assimilation-free CAE-ESN model. First, we validate the framework on the Kuramoto-Sivashinsky equation, in which predictions from the assimilation-free model diverge from the ground truth. Using as few as 25 measurements, the DA-CAE-ESN achieves a negligible  error, which enables long-term forecasts. Robustness is tested across different sampling frequencies, numbers of sensors, and noise levels.
Second, we test the DA-CAE-ESN on the two-dimensional Navier-Stokes equation (Kolmogorov flow) in both quasi-periodic and chaotic regimes. In the quasi-periodic regime, assimilating measurements from 2\% of the grid points in the full flow field enables accurate long-term predictions of the vorticity field and dissipation rate.  In the chaotic regime, the DA-CAE-ESN provides stable forecasts with correct long-term statistics, whereas the assimilation-free CAE-ESN becomes unstable (Fig.~\ref{fig:CAEESN_vs_ENKF_dissipation_34}). The robustness of the DA-CAE-ESN is tested against three sensor placement strategies: uniformly random sampling, equidistant placement on a regular grid, and proper orthogonal decomposition. All three sensor placements yield small errors. 
The augmented state assimilation acts as a localization strategy, which mitigates spurious correlations that arise in high-dimensional systems. The proposed framework can be employed in other autoencoder-based ROMs, e.g., with  transformers and recurrent neural networks, which opens opportunities for real-time prediction of chaotic systems. 

\section*{Data Availability Statement}
The Kolmogorov flow data is available upon request or can be generated using \href{https://github.com/google/jax-cfd}{google/jax-cfd}. The KS data and DA-CAE-ESN are available on GitHub at \href{https://github.com/MagriLab/DA-CAE-ESN}{MagriLab/DA-CAE-ESN}.

\section*{Acknowledgements}
The authors thank Defne Ege Ozan for insightful discussions regarding the ESN-EnKF.

\section*{Funding sources}
 This research has received financial support from the ERC Starting Grant No. PhyCo 949388 and the EPSRC Grant No. EP/W026686/1.

\bibliographystyle{elsarticle-num} 
\bibliography{reference}

\appendix
\section{CAE-ESN Training details}\label{app:cae-esn-details}

The architecture details of the CAE and the multi-scale CAE used to compress the KS and Kolmogorov systems into a manifold are listed in Tab.~\ref{tab:CAE-details}. 
All autoencoders were implemented in 
\href{https://pytorch.org/}{\texttt{PyTorch}} and trained on a single NVIDIA RTX 8000. A tutorial for the CAE-ESN implementation is available on  GitHub (\href{https://github.com/MagriLab/LatentStability}{\texttt{MagriLab/LatentStability}}). 
\begin{table}
\begin{center}
    
\caption{
Design details of the CAE for the  
Kuramoto-Sivashinsky equation and the Multi-Scale CAE for the 
Kolmogorov flow with 3 parallel CAEs.}\label{tab:CAE-details}
\begin{tabular}{lcc}
\toprule
 & KS  (\S\ref{sec:ks-results}) & Kolmogorov (\S\ref{sec:results-kolmogorov})\\ \toprule 
Input dimension & $N=128$ & $N=48 \times 48$ \\ \hline 
Encoder layers & 
\begin{tabular}{r}4 Conv1D \\ 1 Dense\end{tabular} &
\begin{tabular}{r}4 Conv1D \\ 1 Dense\end{tabular}$\left.\begin{matrix} \\[1em]\end{matrix}\right\} $ per CAE \\ \hline 
Decoder layers & 
\begin{tabular}{r} 3 ConvTranspose1D \\ 3 ConvTranspose1D\end{tabular}
&
\begin{tabular}{r} 1 Dense \\
4 ConvTranspose2D\end{tabular}$\left.\begin{matrix} \\[1em]\end{matrix}\right\} $ per CAE\\ \hline 
Activation & $\tanh$ & $\tanh$ \\ \hline 
Initialization & He/Kaiming~\cite{he2015delving} & He/Kaiming~\cite{he2015delving}\\ \hline 
Channels & $1 \to 2 \to 4 \to 8 \to 16$ & $1 \to 4 \to 8 \to 16 \to 32$ \\ \hline 
Kernel sizes & $8 \to 8 \to 3 \to 3$ & $3\times3, 5\times5, 7\times7$ \\ \hline 
Strides & $1 \to 2 \to 2 \to 2$ & $1 \to 2 \to 2 \to 2$ \\
\bottomrule
\end{tabular}
\end{center}
\end{table}
We select the ESN hyperparameters $\{\rho, \sigma_\text{in}, \beta\}$ during validation.  First, we perform a grid search within the ranges of Tab.~\ref{table:esn_parameters} and evaluate the ESN on the validation data. Next, we employ Bayesian optimization to find hyperparameters which minimize the validation loss~\cite{racca2021robust}. 
\begin{table}
\centering
\caption{Parameters of the ESN, where $^*$ indicates optimization range.}\label{table:esn_parameters} 
\begin{tabular}{ lll } 
  \hline
  Spectral radius &$\rho^*$ &   $ \left[0.1, 1.25\right]$   \\ 
  \hline 
   Input scaling &$\sigma_\text{in}^*$ &  $[ \left[0.01, 10.0\right]$\\ 
  \hline 
   Tikhonov regularization & $\beta^*$&  $ \left[10^{-3}, 10^{-11}\right]$\\
  \hline
    Connectivity& $d$ & $10$ \\ 
  \hline
    Input bias &$b_\text{in}$ &  $1$ \\ \hline
\end{tabular}
\end{table}




\end{document}